\documentclass[12pt]{article}

\def\be{\begin{equation}}
\def\ee{\end{equation}}
\def\bse{\begin{subequation}}
\def\ese{\end{subequation}}
\def\ba{\begin{eqnarray}}
\def\ea{\end{eqnarray}}

\usepackage{color,graphicx}
\usepackage{epstopdf,fullpage}
\usepackage{tabularx}
\usepackage{amsmath,amssymb}
\usepackage{hyperref}

\newcommand{\degree}{{}^o}

\usepackage{natbib}
\newcommand{\R}{\mathbb{R}}
\newcommand{\Z}{\mathbb{Z}}
\newcommand{\T}{\mathbb{T}}

\begin{document}

\title{Chaotic and non-chaotic response to quasiperiodic forcing: limits to predictability of ice ages paced by Milankovitch forcing}

\author{Peter Ashwin\\
Department of Mathematics\\
University of Exeter\\
Exeter EX4 4QF, UK
\and
Charles David Camp\\
Mathematics Department\\
California Polytechnic State University\\
San Luis Obispo CA 93407, USA
\and
Anna S. von der Heydt\\
Institute for Marine and Atmospheric Research\\
Department of Physics\\
Utrecht University, Princetonplein 5\\
3584 CC Utrecht, The Netherlands
}

\maketitle

\begin{abstract}
It is well known that periodic forcing of a nonlinear system, even of a two-dimensional autonomous system, can produce chaotic responses with sensitive dependence on initial conditions if the forcing induces sufficient stretching and folding of the phase space. Quasiperiodic forcing can similarly produce chaotic responses, where the transition to chaos on changing a parameter can bring the system into regions of strange non-chaotic behaviour. Although it is generally acknowledged that the timings of Pleistocene ice ages are at least partly due to Milankovitch forcing (which may be approximated as quasiperiodic, with energy concentrated near a small number of frequencies), the precise details of what can be inferred about the timings of glaciations and deglaciations from the forcing is still unclear. In this paper, we perform a quantitative comparison of the response of several low-order nonlinear conceptual models for these ice ages to various types of quasiperiodic forcing.  By computing largest Lyapunov exponents and  mean periods, we demonstrate that many models can have a chaotic response to quasiperiodic forcing for a range of forcing amplitudes, even though some of the simplest conceptual models do not. These results suggest that pacing of ice ages to forcing may have only limited determinism. 
\end{abstract}



\section{Introduction}
\label{sec:intro}

Periodically forced nonlinear oscillators are textbook examples of nonlinear systems whose attractors can exhibit chaotic behaviour \citep{Ott:1993}. Chaotic attractors can similarly be found for more complex (e.g. quasiperiodic) forcing \citep{Yagasaki:1994} along with other structures such as strange nonchaotic attractors \citep{FeudelKuznetsovPikovsky:2006,LaiFeudelGrebogi:1996}.

Since the work of Milankovitch \citep{Milankovitch:1941}, it has been suggested that forcing of the earth's climate by slow, approximately quasiperiodic, changes in the earth's orbit have been responsible for repeated periods of slow growth (glaciation) and rapid retreat (deglaciation) of the Northern Hemisphere ice sheets over the last few million years. These orbital variations, in turn, create oscillations in the northern-hemisphere summer insolation on time scales of obliquity (about 41~kyr) and precession 
(several periodicities between 19 and 23~kyr, \cite{Berger:1978fe}).
The climatic precession cycles are modulated by eccentricity (100~kyr and 400~kyr) \citep{BergerLoutre:1991,Laskaretal:2004}. The  observed glacial cycles of the Pleistocene transition from rather weak ice ages cycling on a 40~kyr time scale during the early Pleistocene ($\sim$~2.5~Myr -- 1~Myr ago) to large-amplitude, asymmetric ice ages cycling on a 100~kyr time scale during the Late Pleistocene ($\sim$~1~Myr ago -- present) \citep{Huybers:2007,LisieckiRaymo:2005}.
In an influential paper, Hays, Imbrie and Shackleton~\citep{Haysetal:1976} suggest that the summer insolation forcing is (i) the fundamental cause of the observed oscillations in glacial cycles of the late Pleistocene and (ii) important for pacing these ice ages (in particular, the timing of the deglaciations).

However, since its introduction there have been a number of problems with this hypothesis, particularly for the late-Pleistocene glacial cycles \citep{Imbrie:1980um,Ruddiman:1986,Imbrie:1993ij,Lisiecki:2010,Paillard:2015is}. 
(i) During the entire Pleistocene, the astronomical forcing is dominated by variations on the precession (about 23~kyr) and obliquity (about 41~kyr)  time scales, but the late Pleistocene response is dominated by variations on the 100~kyr time scale. (ii) The relatively weak forcing by eccentricity on the 100~kyr time scale is decreasing in power throughout the mid and late Pleistocene while the power of the 100~kyr response is growing. (iii) The power in variations of the eccentricity forcing on the 400~kyr time scale is comparable to the power at the 100~kyr time scale, but there is negligible power in the glacial cycle response around 400~kyr. None of results are consistent with the hypothesis that the late-Pleistocene glacial cycles are a direct response to astronomical forcing.

On the other hand, there is ample evidence that the timing of the late-Pleistocene glacial cycles are influenced by the phase of the obliquity and/or eccentricity forcing \citep{Huybers:2007,Lisiecki:2010}. 
Feedback processes internal to the climate system, for example affecting the ice mass 
balance, could then amplify the response to astronomical forcing on specific time scales such that the 
100~kyr time scale appears in the late Pleistocene climate response \citep{AbeOuchietal:2013}. Another 
possibility is that the emergence of the long late-Pleistocene cycles is related to an internal oscillation in 
the climate system, with approximate period of 100~kyr or longer 
\citep{Ditlevsen2009,Crucifix2012a,Daruka:2016hx}, excited by the Milankovitch forcing starting 
around the Mid-Pleistocene transition. Such an internal oscillation has its own periodicity, but if it is 
paced by the quasi-periodic astronomical forcing the period observed in the final response can be 
different from the unforced case. In this paper we focus on the modification of an internal oscillation by quasi-periodic forcing.

The word {\em pacing} suggests a causal but indirect link between a complex multi-period forcing and a response that is somewhat weaker than synchronization. This relates to the key questions around the predictability of future glaciations and deglaciations. This paper aims to further explore the connection between quasiperiodic forcing of nonlinear oscillators and the predictability of the Pleistocene ice ages. We have been inspired by work of Crucifix and collaborators \citep{Crucifix2012a,DeSaedeleer:2013dk,Mitsui:2015hea} to explore several of the oscillator models in the literature under periodic and quasiperiodic forcing. We note that there are however other types of nonlinear resonance, such as response to a resonance system that has no natural oscillations in the absence of a varying input \citep{Marchionne:2016vo}.  

There is a variety of low-order models for the late Pleistocene ice ages, ranging from simple (nonlinear) oscillators to more physically motivated models. For example the ubiquitous van der Pol oscillator has been used to describe general features of the ice-age dynamics and has been suggested as a possible minimal model \citep{DeSaedeleer:2013dk}. Some of these models such as the van der Pol oscillator are strongly constrained: chaotic attractors are associated with canard behaviour where parts of the attractor are contained in strongly repelling parts of phase space \citep{ItohMurakami:1994}. In other words, a forced van der Pol oscillator only gives chaotic attractors for very thin regions in parameter space \citep{Boldetal:2003}. This is however not a universal property of forced nonlinear oscillators - they may have chaotic attractors for much larger regions in parameter space.

For the remainder of Section~\ref{sec:intro}, we briefly review nonlinear oscillator models of the Pleistocene ice ages and possible roles of synchronization and chaos in understanding pacing. In Section~\ref{sec:2dmodels}, we consider the van der Pol oscillator and note that a slightly more complex model (in particular, the van der Pol-Duffing oscillator) can have much wider parameter regimes of stable chaos than seen in \citep{DeSaedeleer:2013dk}. Section~\ref{sec:threedim} studies two low order physics-based models of Saltzmann-Maasch 1991, Palliard-Parrenin 2004: we quite readily find attracting chaotic responses of these models to quasiperiodic forcing. Finally, in Section~\ref{sec:discuss}, we discuss implications of the results and some open problems. Although our studies do not imply the response of the Pleistocene ice-ages to astronomical forcing is necessarily chaotic, one cannot rule out this possibility. We consider what this would imply about the determinism of deglaciations over long timescales. Some computational details are in Appendix~\ref{sec:appendix}.

\subsection{Nonlinear oscillator models of the Pleistocene ice ages}

Many different mechanisms have been suggested, both in terms of mathematical structure and physical basis, for the remarkable late Pleistocene ice ages. For example, \cite{Hagelberg:1994ix,Imbrie:1993ij} suggest a (linear) resonance with the eccentricity forcing could explain the 100~kyr periodicity of the late Pleistocene ice ages and at the same time the absence of a 400~kyr periodicity in the records. However, other analyses of proxy records and models seem to suggest they are governed by a nonlinear process that is in some sense phase-locked to astronomical forcing \citep{Tziperman:2006he}. This mechanism relies on the precession and obliquity time scales of Milankovitch-forcing and requires a nonlinear oscillator whose period varies with amplitude. A nonlinear resonance condition determines the glacial period, which according to the phase-locking mechanism should be `quantized' (meaning that the ratio of the oscillation period and one forcing period should be a ratio of two integers), although this depends on the definition of a `phase' which is not obvious for the physical system or the data. 

As discussed in \cite{Crucifix2012a,DeSaedeleer:2013dk,Mitsui:2015hea}, conceptual models play an important role in terms of providing terminology and mechanisms that might be understood as pacing. It is presumably a sign of usefulness of these ideas that many different nonlinear oscillator models have been found that can produce time series that match the ice volume record. However, it is not even clear whether the glacial cycles should be self-sustaining (existing without forcing) or have stable states that are excited by the forcing. \cite{Tziperman:2006he} suggest the ice-age problem can be usefully be split into two sub-problems, where one involves the understanding of the pacing and timing of glacial cycles while the other should seek for a physical mechanism giving rise to the glacial cycles. The nonlinear phase locking mechanism provides a potential explanation for the first sub-problem, but makes the second more difficult because models with different mechanisms can equally well match the proxy records - this paper concentrates on the first sub-problem.

The models we consider in this paper have the form
\be
\frac{d}{dt}x=f(x,\Lambda(t))
\label{e:original}
\ee
where some finite-dimensional internal dynamics on $x\in\R^d$ governed by $f(x,\lambda)$ represents processes that model the ice-age dynamics for a fixed insolation anomaly $\lambda$, and $\Lambda(t)$ represents the time-dependent insolation anomaly induced by astronomical forcing. We measure $t$ in units of kyr. The models we consider all have periodic dynamics for the unforced case ($\Lambda=0$ is constant) and show a nonlinear response to forcing. We consider several types of forcing, the first being periodic
\be
\Lambda_0(t)=k_1\sin(\omega_1 t),
\label{e:F0forcing}
\ee
where $\omega_1$ represents pure obliquity forcing, $\omega_1=\frac{2\pi}{41}$. Secondly we consider two-frequency quasiperiodic (QP2) forcing
\be
\Lambda_1(t)=k_1\sin(\omega_1 t)+k_2\sin(\omega_2 t)
\label{e:F1forcing}
\ee
that represent components of the obliquity and precession forcing for $(\omega_1,\omega_2)=(\frac{2\pi}{41},\frac{2\pi}{23})$, where $k_{1,2}$ gives the relative amplitudes of these frequencies (strictly speaking this is quasiperiodic only if $\omega_1/\omega_2$ is irrational). We also consider more realistic multi-frequency quasiperiodicity (QPn) forcing
\be
\Lambda_2(t)=k_1\Lambda_{2,o}(t)+k_2\Lambda_{2,p}(t)
\label{e:F2forcing}
\ee
consisting of the obliquity and precessional components of the summer-solstice insolation forcing at $65^o$N considered in \cite{DeSaedeleer:2013dk}, namely
\begin{align}
\Lambda_{2,o}(t) &= \frac{1}{\mu_1}\sum_{k=1}^{15}\left[s_k \sin(\omega_k t)+c_k \cos(\omega_k t)\right],~~~\Lambda_{2,p}(t) = \frac{1}{\mu_1}\sum_{k=16}^{35}\left[s_k \sin(\omega_k t)+c_k \cos(\omega_k t)\right],
\label{eq:F2oF2p}
\end{align}
where the values of $s_k,c_k,\omega_k$ are listed in \cite[Appendix 1]{DeSaedeleer:2013dk} and we define
$$
\mu_1=\sum_{k=1}^{15} \sqrt{s_k^2+c_k^2}\approx 26.57 ~\mbox{ and }~\mu_2=\sum_{k=16}^{35} \sqrt{s_k^2+c_k^2}\approx 76.93
$$
to provide a comparable normalization of $k_1$ and $k_2$ to (\ref{e:F1forcing}).
This is a good model for the last few million years of insolation, and the peak powers of $\Lambda_{2,o}$ and $\Lambda_{2,p}$ are around $41$kyr and $23$kyr respectively. Note that for the latitude $65^o$N, the precessional components average out over the year, whilst the obliquity components do not. Hence we suggest it is reasonable to vary these components independently as a model for the effects of seasonal integration, even though many authors (e.g. \cite{DeSaedeleer:2013dk}) consider only the case $k_1=k\mu_1$ and $k_2=k\mu_2$. In summary, the QPn forcing (\ref{eq:F2oF2p}) is a more realistic version of the simplified QP2 forcing (\ref{e:F2forcing}). 

We mainly consider four models in our numerical exploration: the 2D models of van der Pol and van der Pol-Duffing, and the 3D models of \cite{Saltzman:1991jl} and \cite{Paillard:2004dn}. All have been proposed as possible models for the Pleistocene ice ages but we highlight the van der Pol model stands out as being dynamically simpler than the others, in that chaos is confined to very narrow regions.

\subsection{Synchronization and phase locking}

In this paper, we try to distinguish between different types of nonlinear models in terms of the response to quasi-periodic forcing. There is a trade-off between simplicity of models for ice age pacing and  realism in terms of physical processes and resolution included.  While there have 
 been attempts to integrate earth system models of intermediate complexity including ocean, 
 ice sheet and carbon cycle dynamics over several glacial cycles 
 \citep{AbeOuchietal:2013,Ganopolski:2017gj}, these models are by far too complex to analyse and 
 understand the process of pacing of the glacial cycles by the astronomical forcing. These questions about pacing can be more adequately answered by studying less complex, simplified models that have been used in the past for the glacial-interglacial cycles. Some of those are physically motivated by including only a limited number of physical processes and restrict the spatial resolution typically to a few ``boxes'' (e.g. \cite{Gildor2001a}), or only one global signal (e.g. \cite{Saltzman:1991jl,Paillard:1998bn}), while others include more mathematical concepts of self-sustained oscillators \cite{Crucifix2012a}.

The strongest form of pacing one can imagine consists of generalized synchronization to the signal, i.e. there is a functional relationship between the instantaneous value of the forcing to that of the signal - this is clearly not a realistic model for signals such as the proxy of \cite{LisieckiRaymo:2005} so we should consider weaker versions. This includes, for example, synchronization to a filtered component of the forcing, or a more loose phase locking such that some phase extracted from the forcing remains close to a phase extracted from the response.

The nonlinear phase-locking mechanism of \cite{Tziperman:2006he} argues that any nonlinear model that becomes appropriately phase-locked would be able to describe the observed ice-volume record.  Consequences of nonlinear phase locking suggested by \cite{Tziperman:2006he} are: (i) Even in the presence of abundant noise in the climate system the phase locking can be effective. (ii) The glacial period should be multiples of the precession and obliquity forcing, although this ratio can change over time. Eccentricity forcing does play a more indirect role by modulating the precession and obliquity forcing amplitudes.  (iii) The quasiperiodic nature of the Milankovitch forcing allows for varying glacial periods; however, it was suggested that the timing of glacial terminations could be uniquely determined (even though the phase of the Milankovitch-forcing is not the same during all glacial terminations but spreads around 1/4 of a cycle in the model used by \cite{Tziperman:2006he}). This last consequence implies the model is insensitive to small changes in initial conditions, because eventually all time series become phase-locked to the same forcing. Similarly, \cite{TzedakisPC:2017kn} use a statistical model to predict the series of past deglaciations suggesting there exists only a small set of possible realisations for the deglaciations during the late Pleistocene. Note that, as highlighted by \cite{Marchionne:2016vo}, complex patterns of nonlinear resonance can appear even in the absence of limit cycle oscillations for the unforced system: we do not consider such cases here.

There is no unique way to extract a continuously varying phase from a complex signal unless it is close to periodic: spectral methods have the disadvantage that they are global in time, assume stationarity of the signal and may have significant end effects for finite time intervals. There are many ways to do this, for example by measuring the rotation around a point for a projection of phase space onto a plane. We use a well-studied method of extracting the phase by using the Hilbert transform of the oscillatory signal $x(t)$ \citep{PikovskyRosenblumKurths:2001,LeVQetal:2001}. Let us denote the mean of $x(t)$ as $\overline{x}$. This phase is
\be
\psi(t)= \arg\left[(x(t)-\overline{x})+ i y(t)\right]
\label{e:hilbertphase}
\ee
where
$$
y(t)=\frac{1}{\pi} ~\mathrm{pv}\int_{s=-\infty}^{\infty} \frac{x(t)-\overline{x}}{t-s}\,ds
$$
is such that $x+ i y$ is analytic: the integral is understood as a Cauchy principal value.  Using this we can compute $R(T)$, the mean number of rotations of the phase of $x(t)$ over the interval $[0,T]$, as 
$$
R(T)=\frac{[\psi(T)-\psi(0)]}{2\pi T}
$$
and the mean period is $1/R(T)$. This works well for signals that have one crossing of the mean $\overline{x}$ per period but is not so reliable for signals with several crossings per period. For a quasiperiodic signal that has many competing frequencies, different linearly filtered signals will show completely different phases and so conclusions from this (as with any) method of phase extraction need to be treated with care.

\subsection{Pacing, attractors and the role of chaos}

At the weaker end of pacing, one could argue there is some sort of synchronization to forcing if the response loses all information about its initial condition - for example if a neighbourhood of initial conditions converge to the same response trajectories. These response trajectories are {\em pullback attractors} for the forced non-autonomous system \citep{DeSaedeleer:2013dk}, and there may be several possible local pullback attractors for a given forcing \citep{Kloeden:2011}.

Given a solution $x(t) \in \R^n$ of (\ref{e:original}), we compute the attracting behaviour for some generic initial condition. For stationary forcing the Lyapunov (characteristic) exponents are given by
$$
\lambda = \lim_{T\rightarrow \infty} \frac{1}{T} \ln \|v(T)\|
$$
where $v(t)$ is a solution of the (in general nonautonomous) ODE
\be
\frac{d}{dt}v=Df(x(t),\Lambda(t))v.
\label{e:variational}
\ee
For a typical initial $(x,v)$, the Lyapunov exponent $\lambda$ can take one of up to $n$ possible values
$$
\lambda_1\geq \lambda_2 \geq \cdots \geq \lambda_n
$$
that represent the possible exponential rates of attraction or repulsion of nearby trajectories: see for example \cite{EckmannRuelle:1985}. The largest Lyapunov exponent (LLE) $\lambda_1$ enables a characterization of the attracting dynamics into chaotic ($\lambda_1>0$) or non-chaotic ($\lambda_1\leq 0$). More details of how the LLE is computed is described in Appendix~\ref{sec:appendix}.

Pullback attractors can be characterised using LLEs of the forced system: they are isolated trajectories in the case of negative LLE while a positive LLE implies the pullback attractor is a chaotic set. Even in the case of QPn forcing and a negative LLE one may find sensitive dependence on phases - this corresponds to {\em strange nonchaotic attractor} (SNA) \citep{LaiFeudelGrebogi:1996} for the system (\ref{e:original}) extended to describe the forcing $\Lambda=L(\theta_1,\ldots,\theta_n)$ where $(\theta_1,\ldots,\theta_n) \in\T^n$ and
$$
\frac{d}{dt} \theta_i=\omega_i~\mbox{ for }i=1,\ldots,n.
$$
For SNAs, there is a unique point pullback attractor for any given realisation of the forcing, but the location of that point may vary wildly on changing one of the phases $\theta_i$ while fixing the others.
SNAs have been extensively studied and found in a variety of QP2 forced systems \citep{FeudelKuznetsovPikovsky:2006}, including phase oscillator models for Pleistocene ice age oscillations \citep{Mitsui:2015hea}.


There is a complex relationship between chaotic behaviour and phase locking - as we will see, QP forcing for many nonlinear oscillators with two or more variables readily leads to chaos in the sense of a positive LLE. The presence of chaos in the response does not however mean there is no phase locking - the chaos may be small scale and cause a jitter of the precise phase with no loss of phase locking. It does imply a form of non-determinism in that the future of the trajectory will depend sensitively on initial condition at all points in the future. However for the models we investigate here, we find chaos that is associated with loss of phase locking.

\section{Simple oscillator responses to forcing}
\label{sec:2dmodels}

We first consider responses of some simple oscillators to periodic and quasiperiodic forcing before moving on to more realistic oscillators in Section~\ref{sec:threedim}.

\subsection{The van der Pol oscillator and generalizations}

The modified van der Pol oscillator was suggested as minimal model for the ice age cycles in \cite{Crucifix2012a,DeSaedeleer:2013dk}. This model has mostly negative LLEs for both periodic and quasiperiodic (two periods) forcing indicating that it unlikely shows a chaotic response to forcing anywhere in the forcing-parameter space. 

The van der Pol model of \cite{Crucifix2012a,DeSaedeleer:2013dk} involves two dependent variables: $x$ is a slow variable representing deviation of ice volume from some reference and $y$ is a fast variable representing some feedback mechanism with hysteresis.
\begin{align}
\tau \kappa\frac{d}{dt} x & = \gamma \Lambda(t)-\beta-y\nonumber\\
\tau \kappa\frac{d}{dt} y & = \alpha(y-y^3/3+x)
\label{e:SCW}
\end{align}
where the forcing term $\Lambda(t)$ represents astronomical forcing and can be either periodic (e.g. only obliquity component) or quasiperiodic. The parameters we consider in Table~\ref{t:VDP} are from \cite{DeSaedeleer:2013dk}, and for convenience and comparability with their results we use two parameters $\tau$ and $\kappa$ to normalise the period around $\tau=1$. System (\ref{e:SCW}) can be written more explicitly as the van der Pol oscillator with a linear restoring force (the nonlinearity is only in the damping):
$$
\frac{\tau^2\kappa^2}{\alpha} \frac{d^2}{dt^2} y+\tau \kappa (y^2-1)\frac{d}{dt}y+y=\gamma \Lambda(t)-\beta.
$$

\begin{table}
\centering
\begin{tabular}{llll}
\hline
Parameter & VDP value & VDPD value & Interpretation\\
\hline
$\alpha$ & $11.11$ & $11.11$ & fast/slow timescale separation\\
$\beta$ & $0.25$ & $0.25$ & symmetry breaking\\
$\gamma$ & $0.75$ & $0.75$ & effective forcing amplitude\\
$\kappa$ & $35.09$ & $52.0$ & unforced oscillation timescale\\
$\tau$ & $1$ & $1$ & time scaling\\
$g(y)$ & $y$ & $-y+1.2y^3$ & restoring force\\
\hline
\end{tabular}
\caption{Default parameters used for the van der Pol (\ref{e:SCW}) and van der Pol Duffing (\ref{e:vdPD}) models: all variables are dimensionless, except for time which is in kyr. Note that a nonlinear response has been chosen that give relaxation-like oscillations and $\kappa$ is chosen to ensure the unforced oscillations have period approximately 100~kyr for time scaling $\tau=1$.\label{t:VDP}
}
\end{table}

\begin{figure}
\centering
\includegraphics[width=8cm]{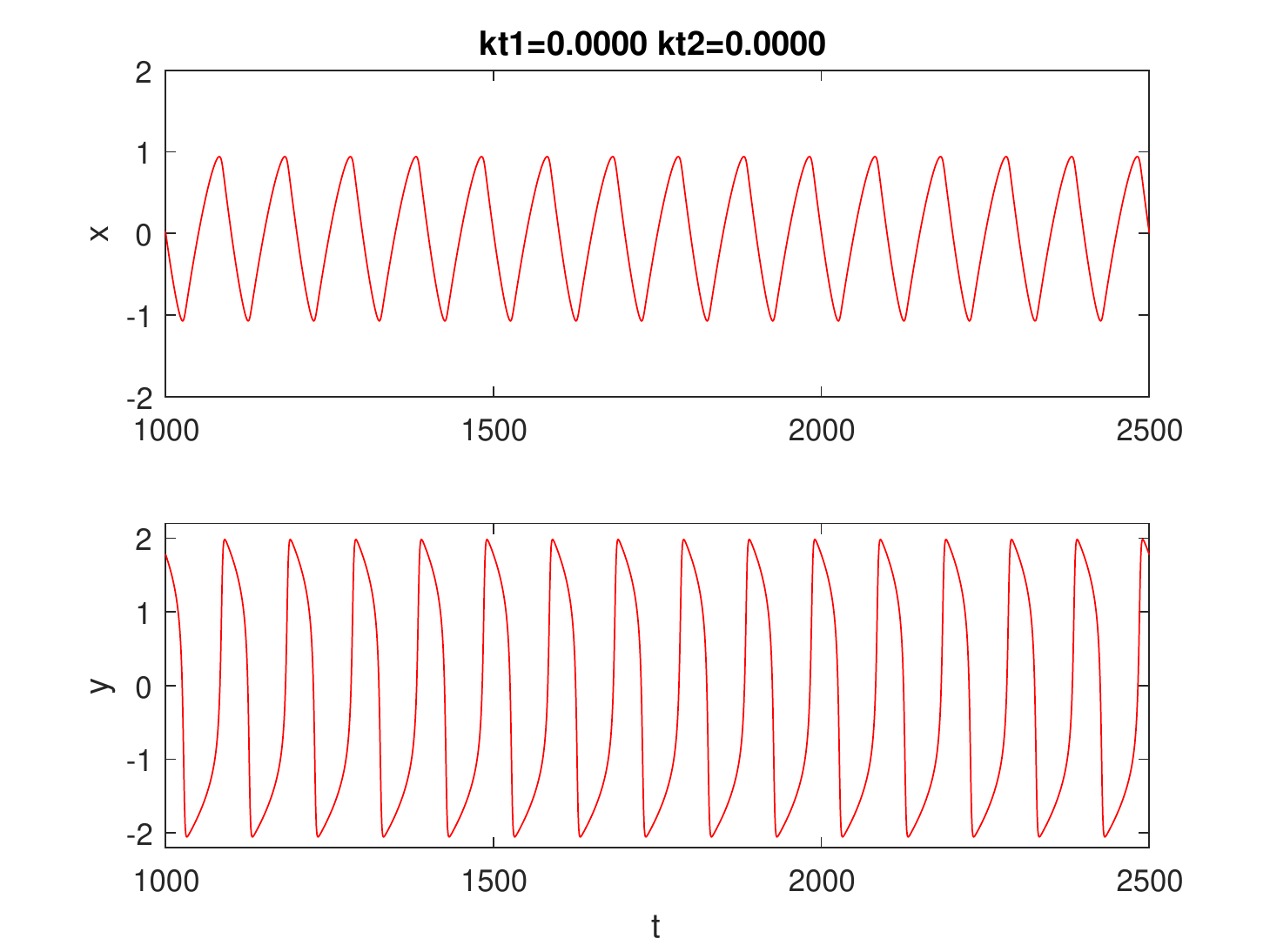}~\includegraphics[width=7cm]{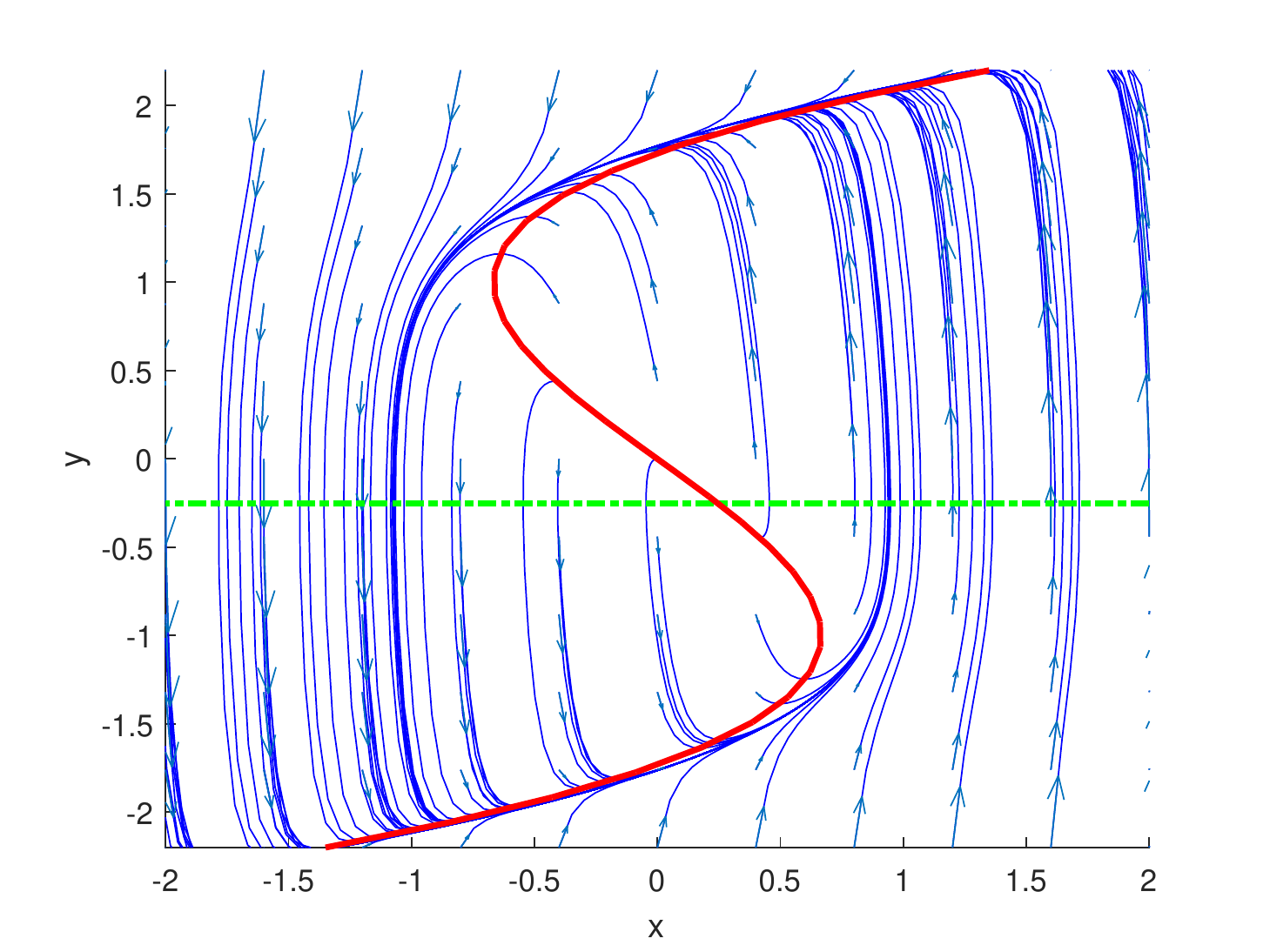}

(a) \hspace{7cm} (b)

\includegraphics[width=8cm]{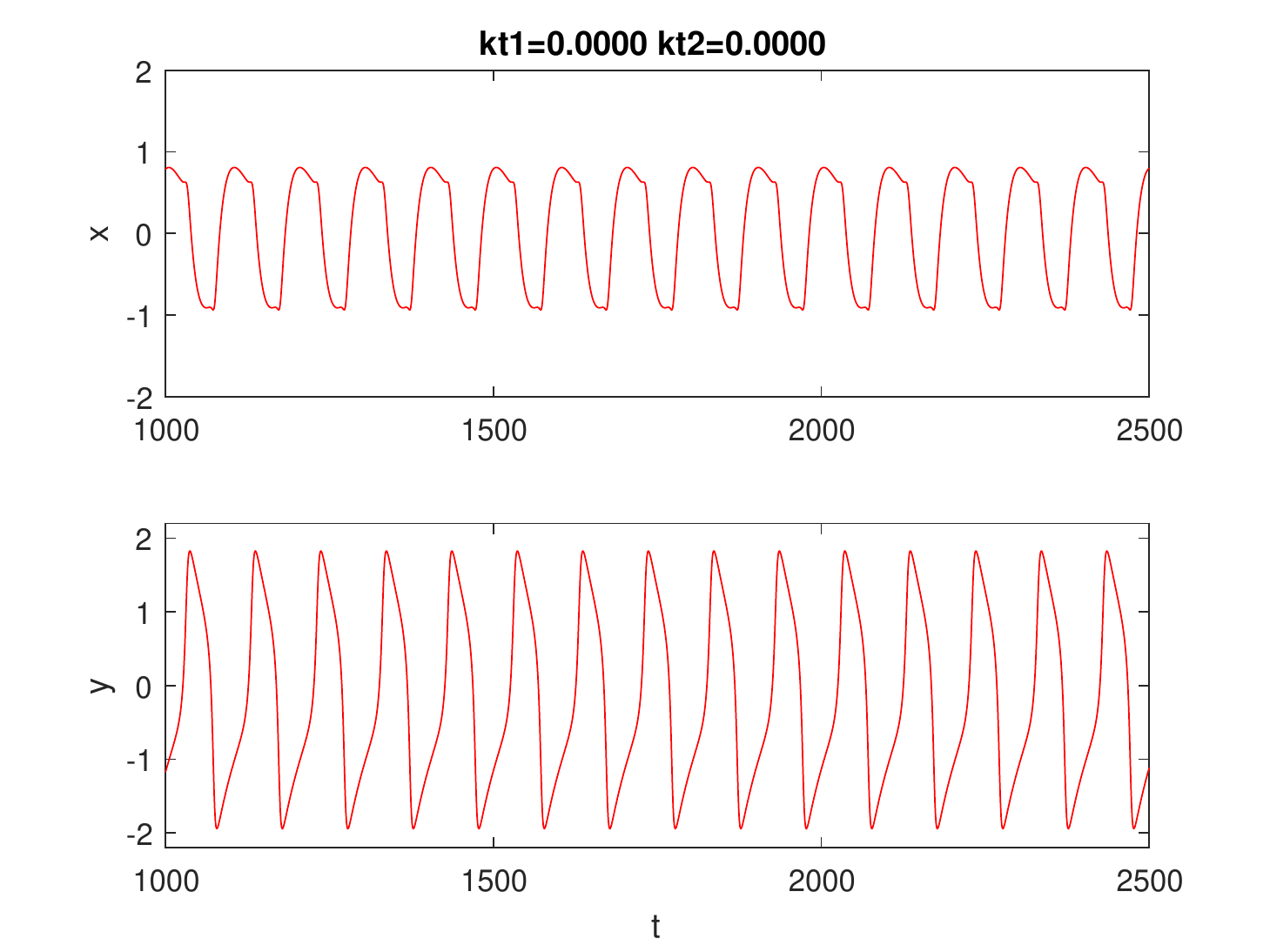}~\includegraphics[width=7cm]{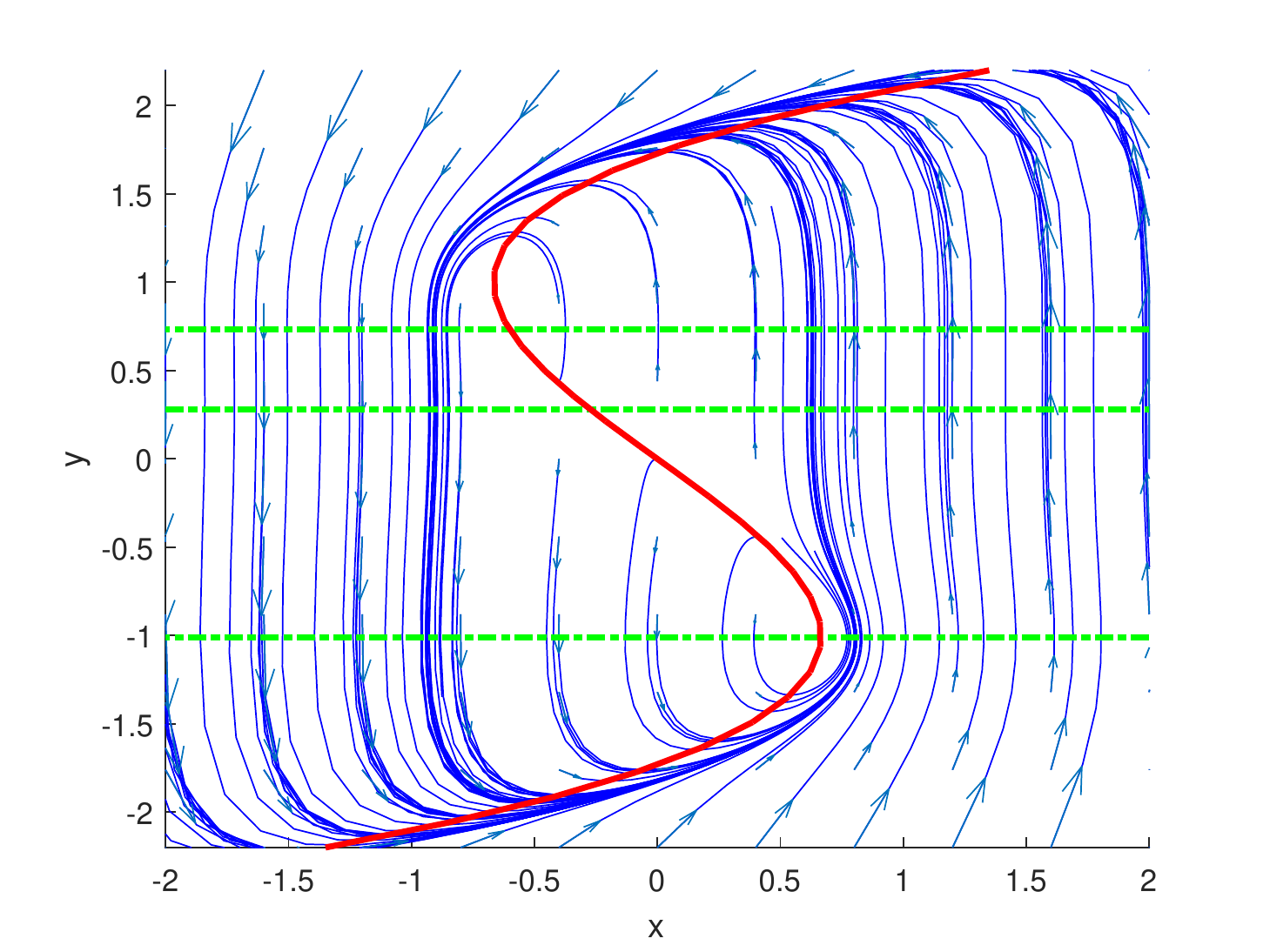}

(c) \hspace{7cm} (d)

\caption{Time series (a,c) and phase portraits (b,d) of the unforced oscillations for the van der Pol (a,b) (\ref{e:SCW}) and van der Pol-Duffing (c,d) (\ref{e:vdPD}) systems, with parameters as in Table~\ref{t:VDP}. In (b,d) the solid red lines show the $y$-nullcline and the dashed green lines shows the $x$-nullcline: note that there are two additional $x$-nullclines for the latter system. Observe that both show relaxation oscillations with approximately the same period. Moreover, the critical manifold (y-nullcline) is identical in both cases. \label{f:unforced_vdpvdpd}}
\label{fig:iceageppvdpqpr20506fig1}
\end{figure}

A generalization of this can be obtained by replacing the linear restoring force with a nonlinear $g(y)$ as in the following. 
\begin{align}
\tau \kappa \frac{d}{dt} x & = \gamma \Lambda(t)-\beta- g(y)\nonumber\\
\tau \kappa \frac{d}{dt} y & = \alpha(y-y^3/3+x).
\label{e:vdPD}
\end{align}
Note that the system (\ref{e:vdPD}) is equivalent to a generalised van der Pol-Duffing oscillator:
$$
\frac{\tau^2\kappa^2}{\alpha} \frac{d^2}{dt^2} y+\tau \kappa (y^2-1)\frac{d}{dt}y+g(y)=\gamma \Lambda(t)-\beta
$$
with nonlinear restoring force $g(y)$. Figure~\ref{f:unforced_vdpvdpd} illustrates typical unforced solutions of the two systems in time series and phase portrait. The van der Pol- Duffing oscillator does show regions of potentially chaotic response to periodic and quasiperiodic forcing depending on the amplitude of the forcing.

\begin{figure}
\centering
\includegraphics[width=8cm]{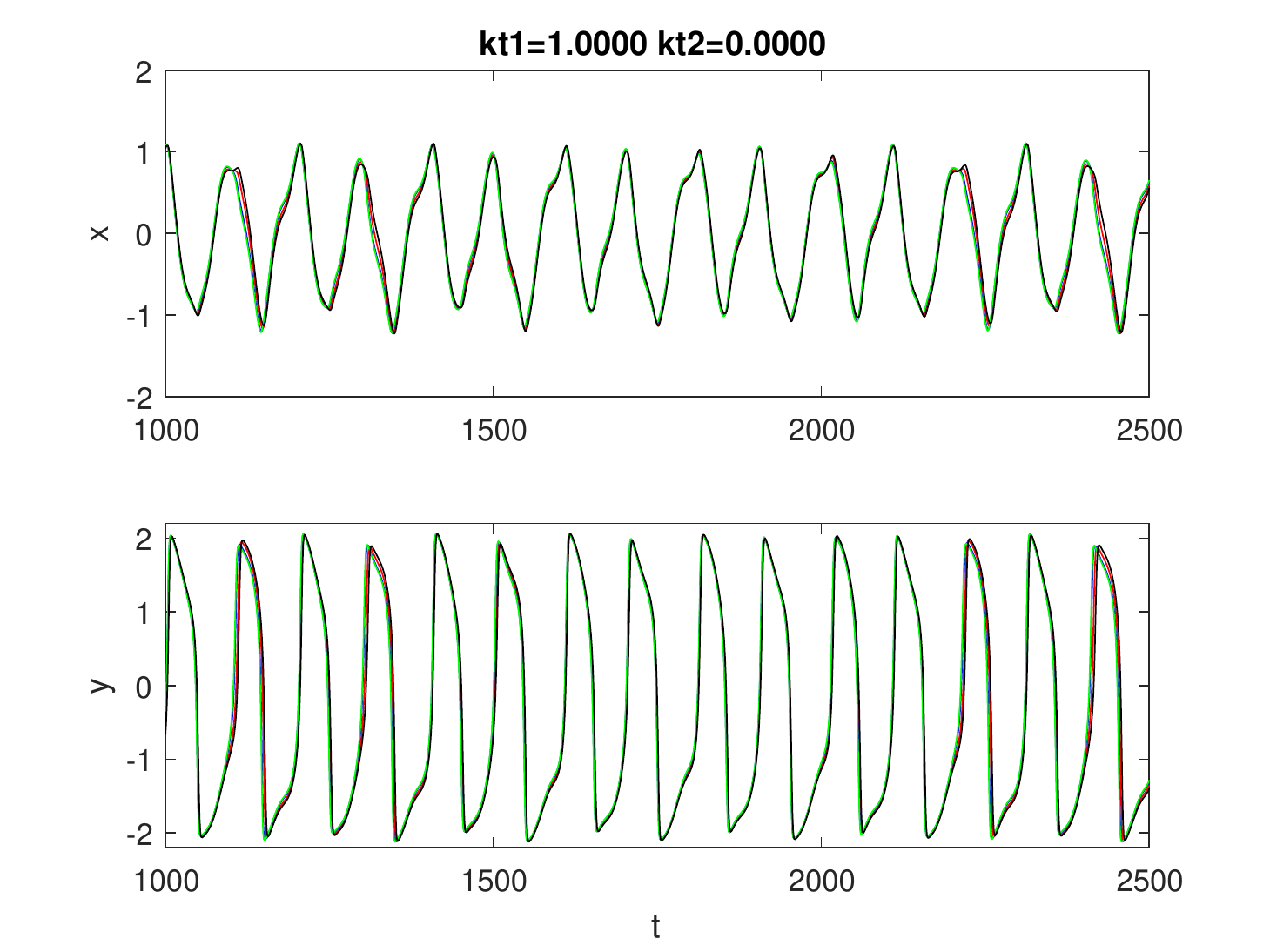}~
\includegraphics[width=8cm]{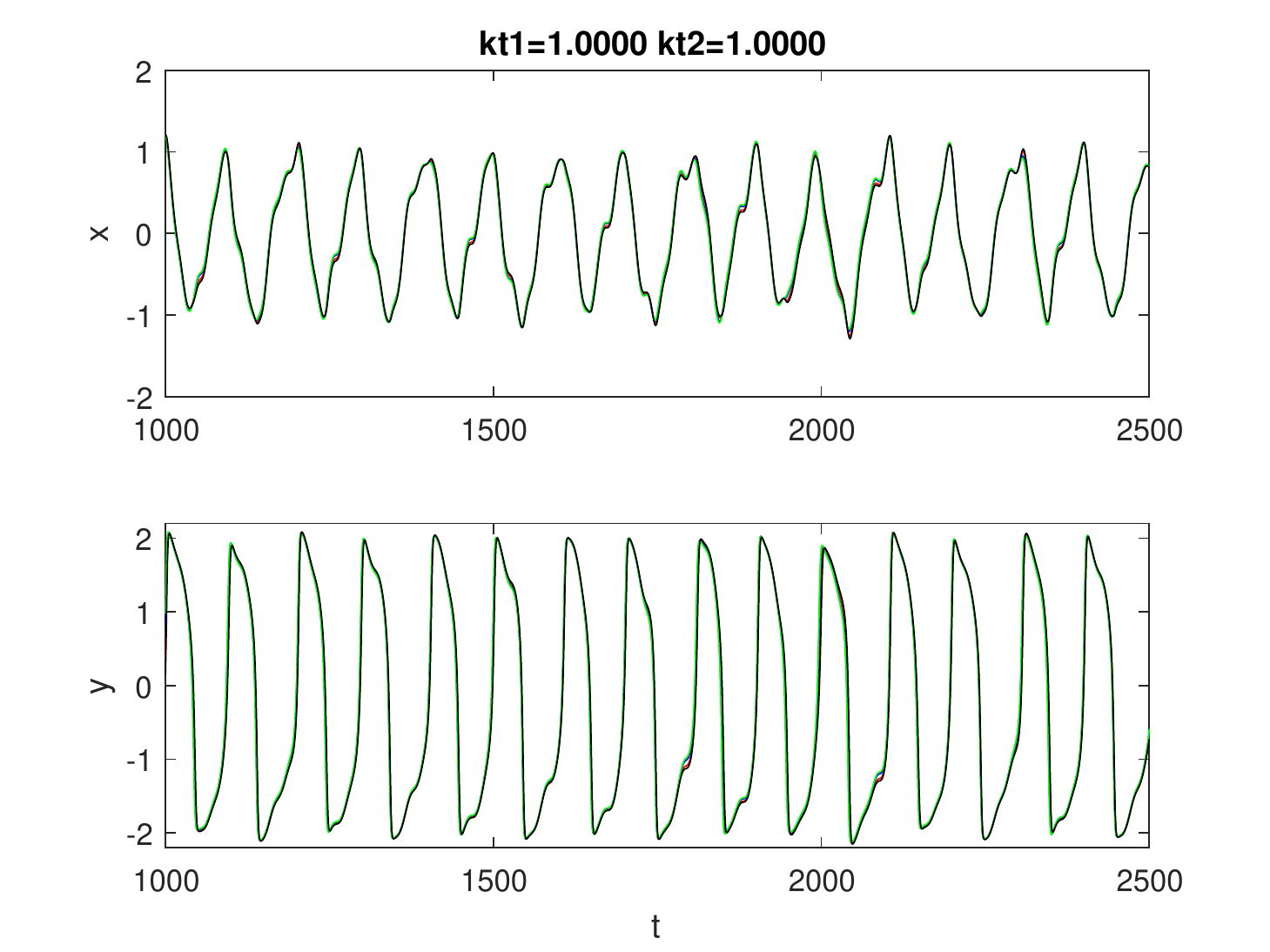}~

(a) \hspace{8cm} (b) 

\includegraphics[width=8cm]{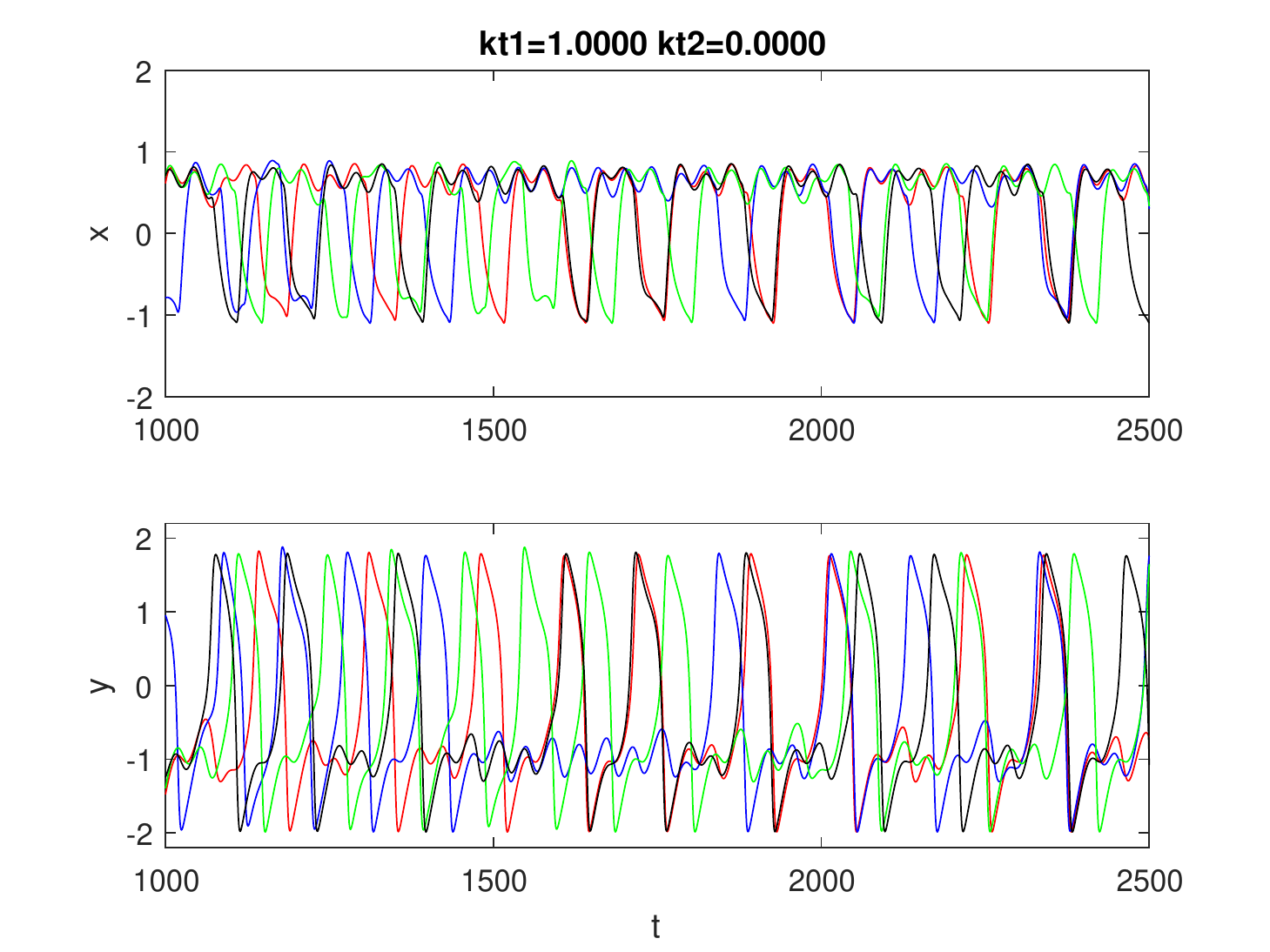}~
\includegraphics[width=8cm]{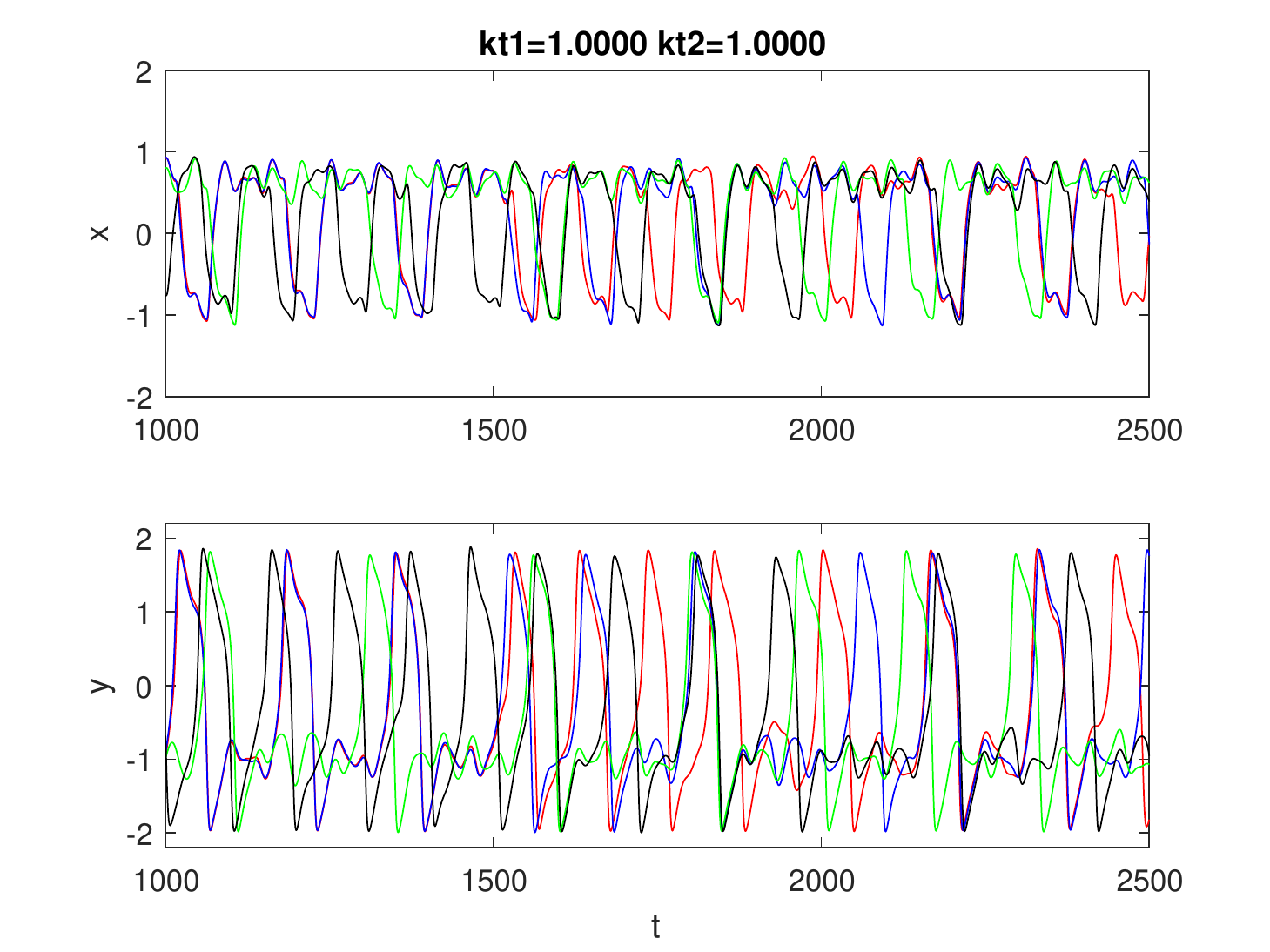}

(c) \hspace{8cm} (d) 

\caption{Time series showing oscillations of the (a,b) van der Pol and (c,d) van der Pol-Duffing system (\ref{e:vdPD}) with typical parameters as in Table~\ref{t:VDP} forced by (\ref{e:F1forcing}) with four random initial conditions started at time $t=0$. For (a,c) there is periodic forcing (\ref{e:F1forcing}) with $k_1=1$, $k_2=0$ while for (b,d) there is QP2 forcing (\ref{e:F1forcing}) with $k_1=1$, $k_2=1$. Note the strong synchronization to a single trajectory for the van der Pol system (a,c) and the presence of intermittent fluctuations in the transitions for the the van der Pol-Duffing system (b,d) typical for a chaotic response.} 
\label{f:forced_vdpvdpd}
\end{figure}

\subsection{Responses to periodic and quasiperiodic forcing}

The relative lack of chaotic behaviour under periodic, QP2 and QPn forcing for the van der Pol oscillator (\ref{e:SCW}) was noted in  \cite{DeSaedeleer:2013dk}. However this is somewhat special - adding nonlinearity in the restoring force can lead to more prevalence of chaos. Figure~\ref{f:forced_vdpvdpd}(a,b) contrasts two examples of timeseries of these oscillators subject to periodic forcing while (c,d) gives a similar contrast for QP2 forcing. For the van der Pol oscillator (Figure~\ref{f:forced_vdpvdpd}(a,c)), four randomly chosen initial conditions synchronize to the same response trajectory; while in the case of the van der Pol-Duffing system (Figure~\ref{f:forced_vdpvdpd}(b,d)), the synchronization is only intermittent.

\begin{figure}
\begin{center}
\includegraphics[width=0.48\textwidth]{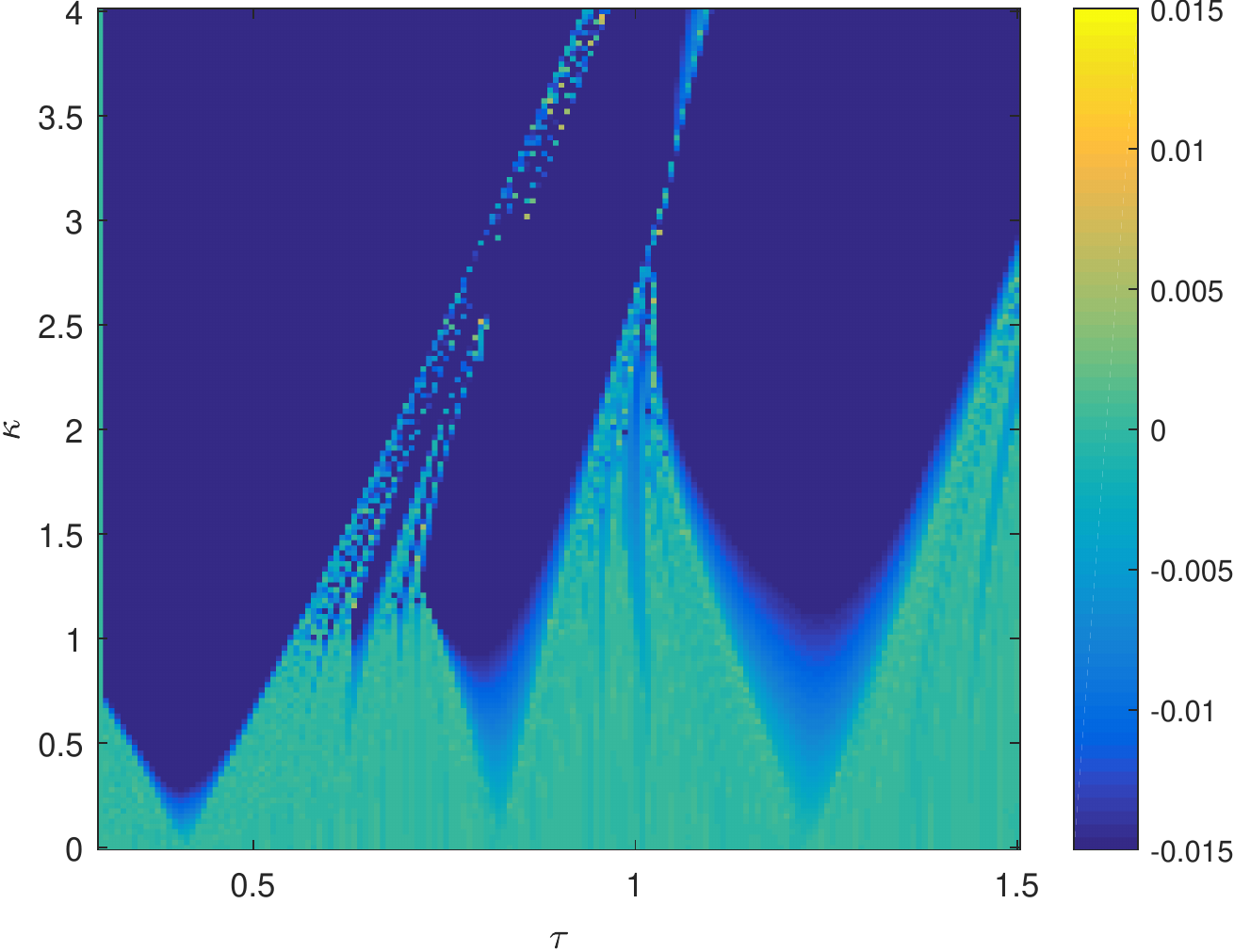}~
\includegraphics[width=0.48\textwidth]{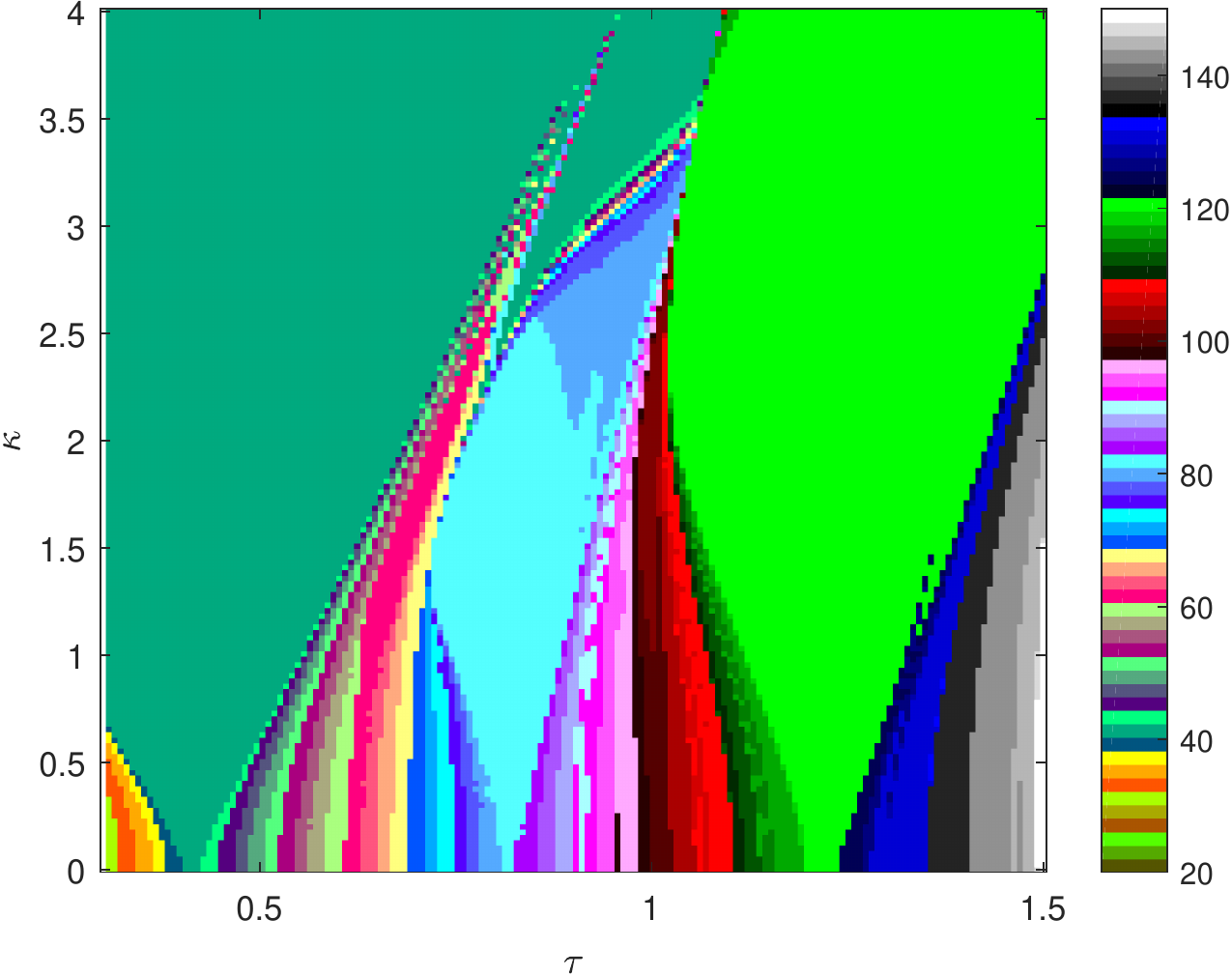}

(a)\hspace{7cm}(b)

\includegraphics[width=0.48\textwidth]{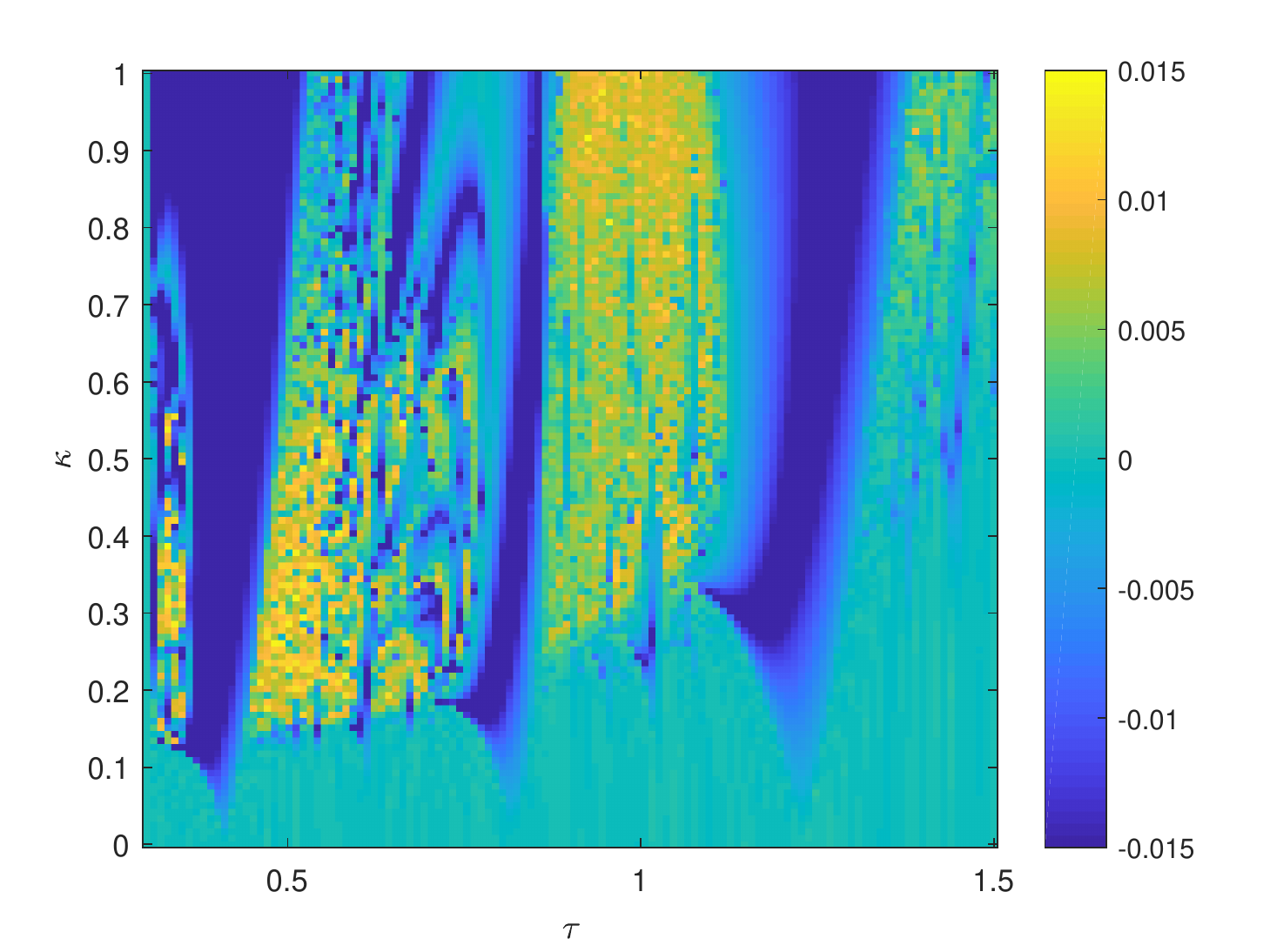}~
\includegraphics[width=0.48\textwidth]{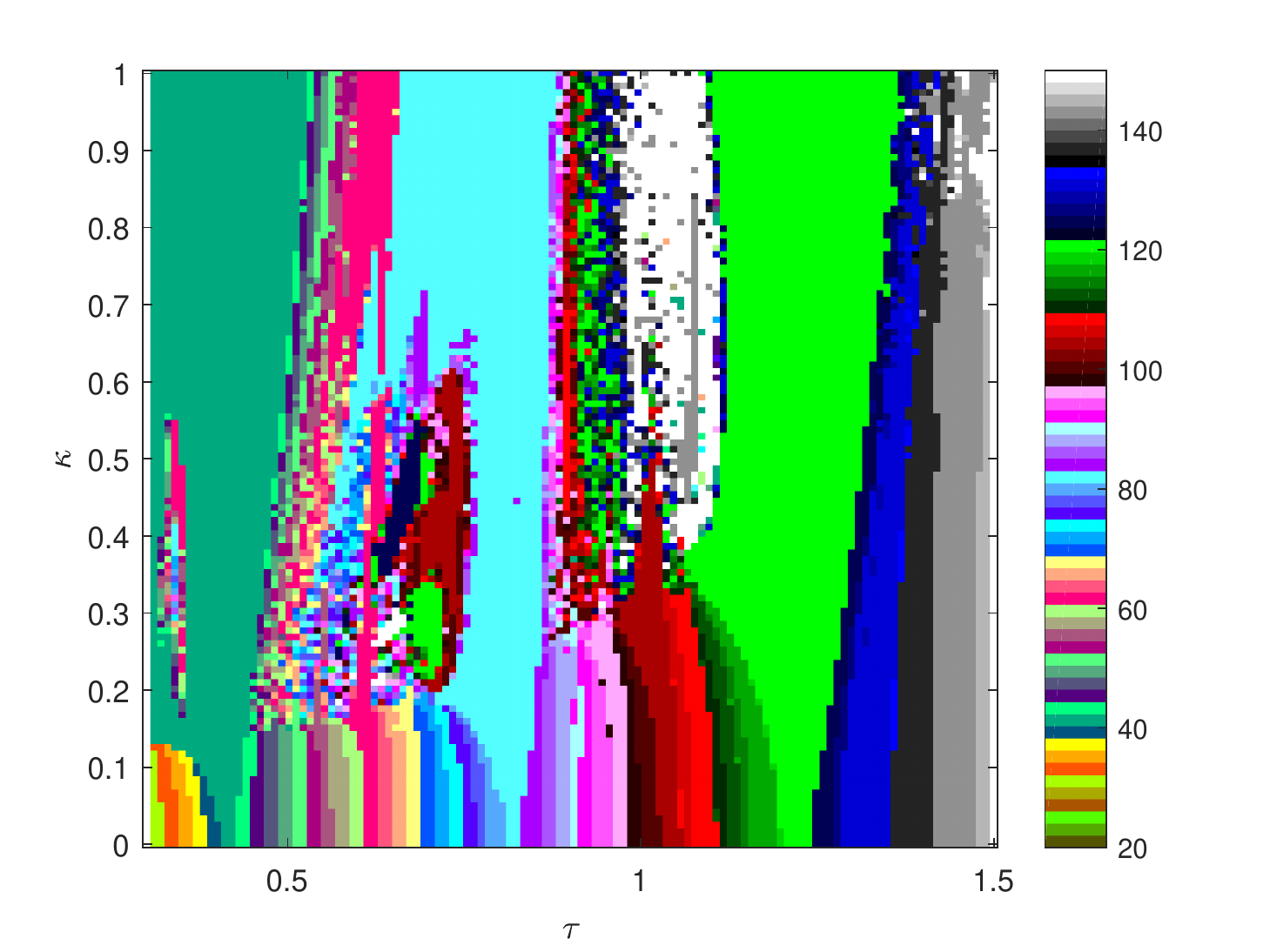}

(c)\hspace{7cm}(d)
\end{center}
\caption{Scan of (a,c) LLEs and (b,d) mean periods for typical initial conditions of (a,b) van der Pol equation (\ref{e:SCW}) and (c,d) van der Pol Duffing equation (\ref{e:vdPD}) with periodic forcing (\ref{e:F0forcing}). We vary the time scaling $\tau$ (where $\tau=1$ corresponds to an unforced period 100~kyr) on the horizontal axis and the forcing amplitude $k=k_1$ on the vertical axis. Observe in both cases there is an Arnold tongue structure for small $k=k_1$ and regions of 1:1, 2:1 and 3:1 locking (left to right) that persist to large $k=k_1$ for (a,b) but include regions of more complex chaotic dynamics for (c,d) (compare with \cite[Figure~6(a,c)]{DeSaedeleer:2013dk}). }
\label{f:vdpdLEscan}
\end{figure}

Scans of the parameter space spanned by the forcing amplitude $k$ and the time scaling $\tau$ in Figure~\ref{f:vdpdLEscan}(a) show the LLE for the van der Pol model (\ref{e:SCW}) with forcing (\ref{e:F0forcing}), while Figure~\ref{f:vdpdLEscan}(b) show the LLE for a similar scale with the van der Pol-Duffing generalization (\ref{e:vdPD}).  Note that $\tau$ varies the unforced period with $\tau=1$ corresponding to the expected 100~kyr in both cases: compare (a,b) to \cite[Fig 6]{DeSaedeleer:2013dk}. As expected, for small forcing amplitudes $k_1$, weak forcing of a limit cycle oscillator gives frequency locking on an attracting invariant torus. These Arnold tongue regions of $p$:$q$ frequency locking appear \citep{PikovskyRosenblumKurths:2001} where the ratio of forcing to natural frequency is close to $p:q$. The tongues where $p$ and $q$ are small grow preferentially with $k_1$, meaning that at for larger $k_1$ in (a,b) we see most of parameter space is taken up by low order frequency locking with ratios $1$:$1$, $2$:$1$ and $3$:$1$. 

For larger $k_1$ these regions start to overlap, associated with break-up of the invariant torus. This torus break up means the attractor can explore a higher dimensional region in phase space and potentially become chaotic. However, chaos is notable by its absence in Figure~\ref{f:vdpdLEscan}(a). In fact, chaotic attractors are forced to live in `cracks' between frequency locking. This sparsity of chaos for the forced van der Pol oscillator has been studied in the literature \citep{Boldetal:2003,ItohMurakami:1994}, and is associated with the fact that chaotic attractors involves canard trajectories  which confine the chaos to thin strips in parameter space for this model. 

By contrast, the forced van der Pol-Duffing oscillator has chaos associated with recurrent approaches to a saddle equilibrium which makes it more robust in parameter space, as illustrated in Figure~\ref{f:vdpdLEscan}(c,d): it does not rely on canard trajectories to stretch and fold neighbourhoods in phase space. 

\begin{figure}
\begin{center}
\includegraphics[width=0.48\textwidth]{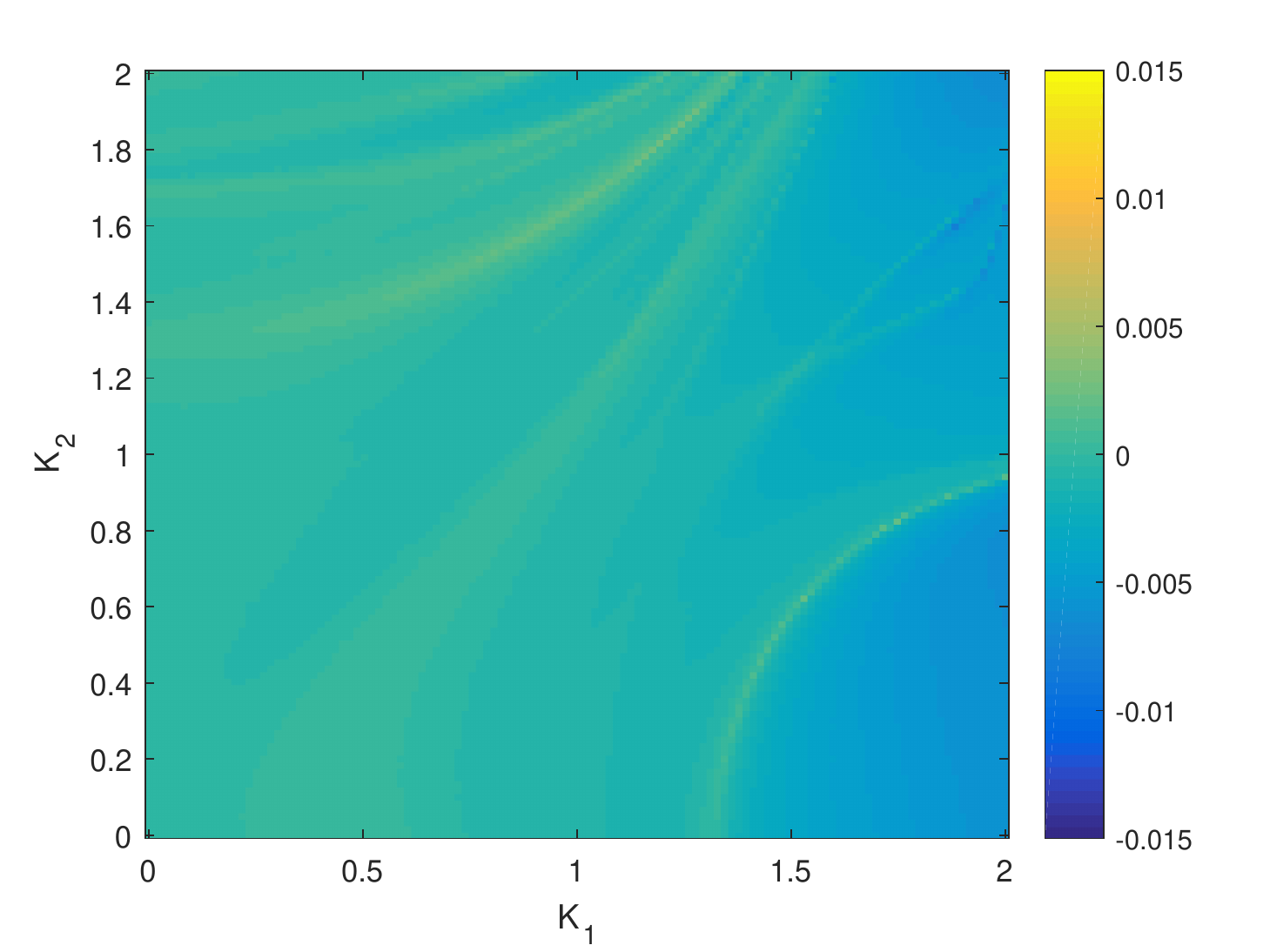}~
\includegraphics[width=0.48\textwidth]{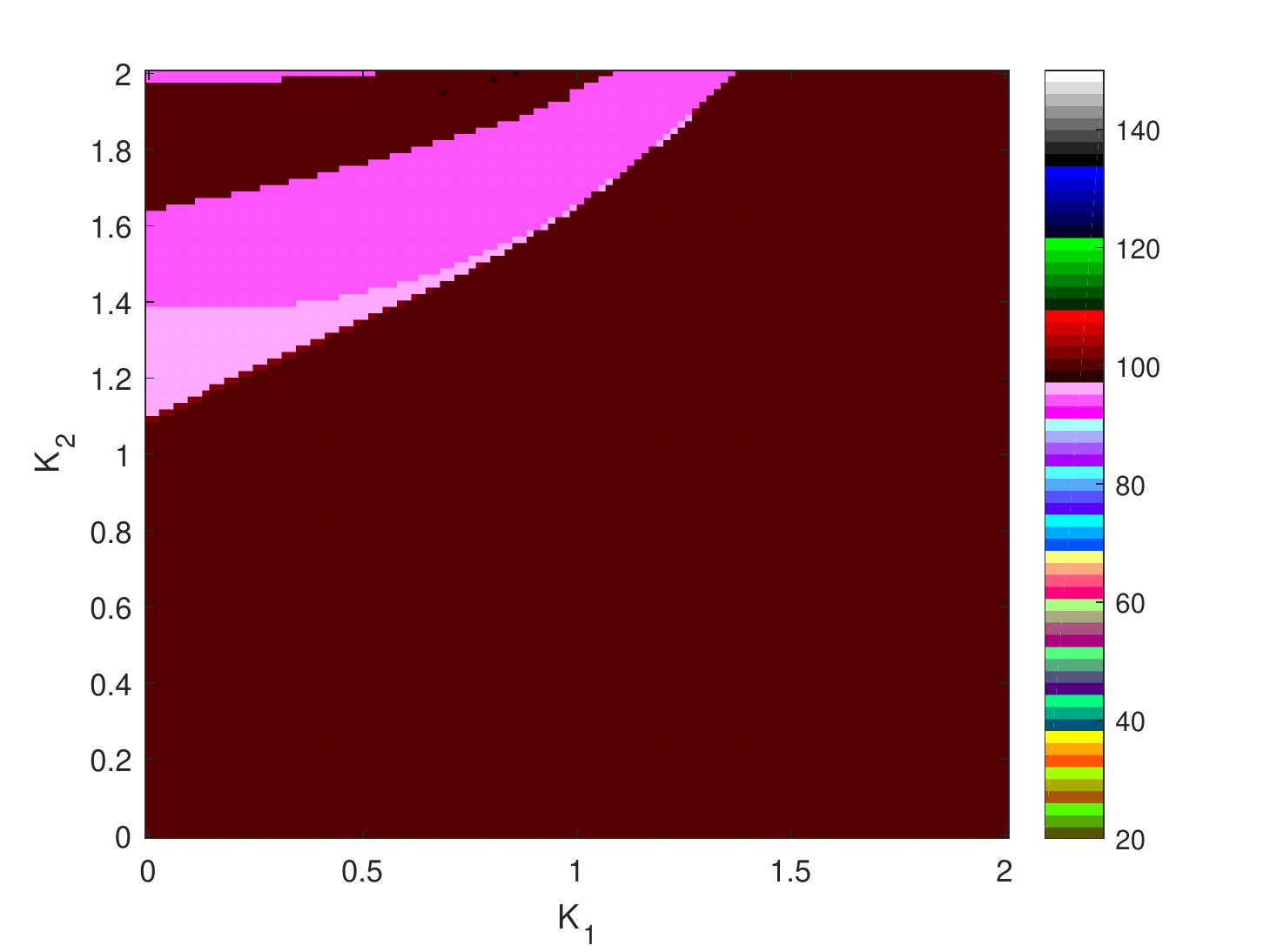}

(a)\hspace{7cm}(b)

\includegraphics[width=0.48\textwidth]{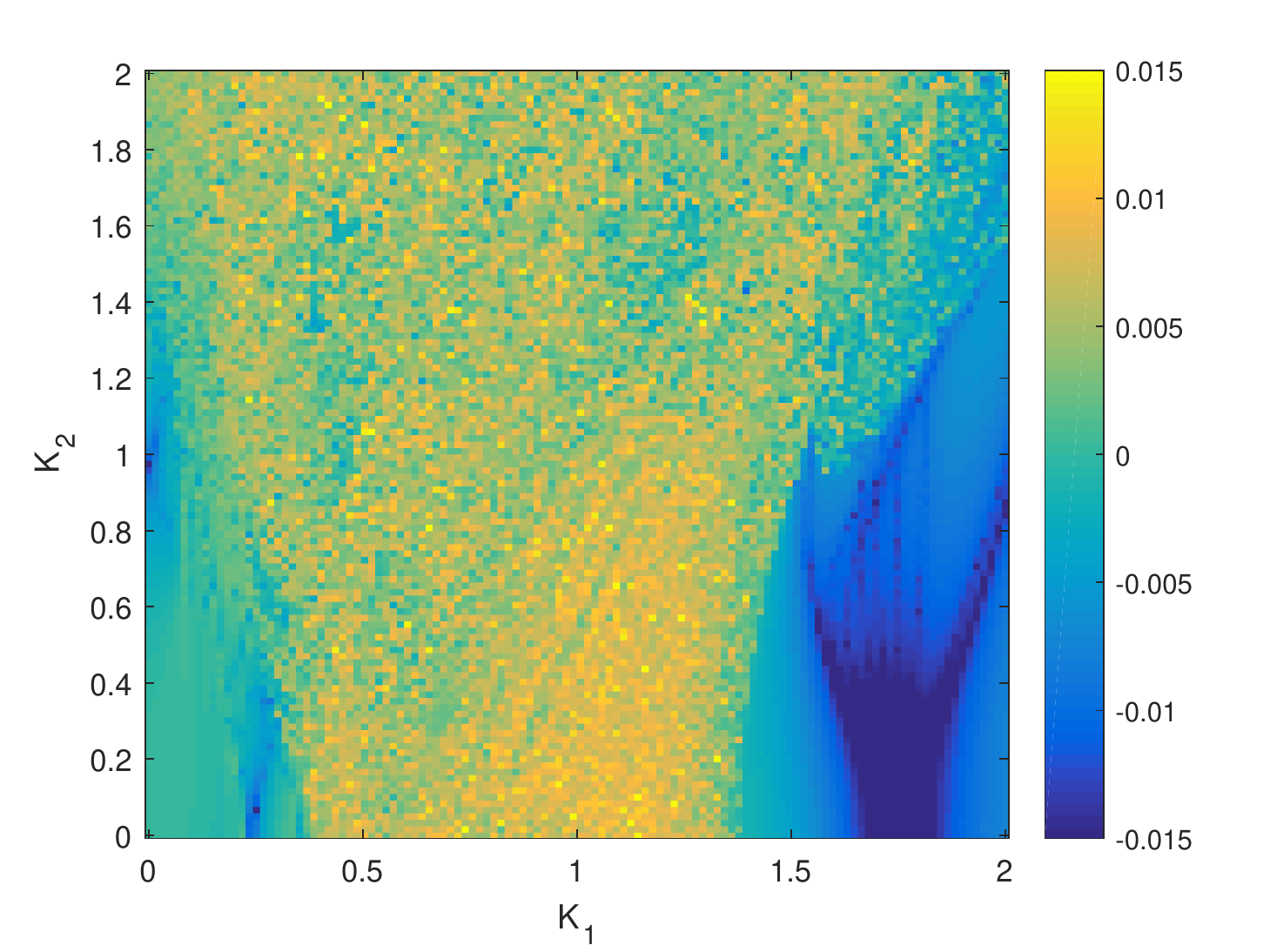}~
\includegraphics[width=0.48\textwidth]{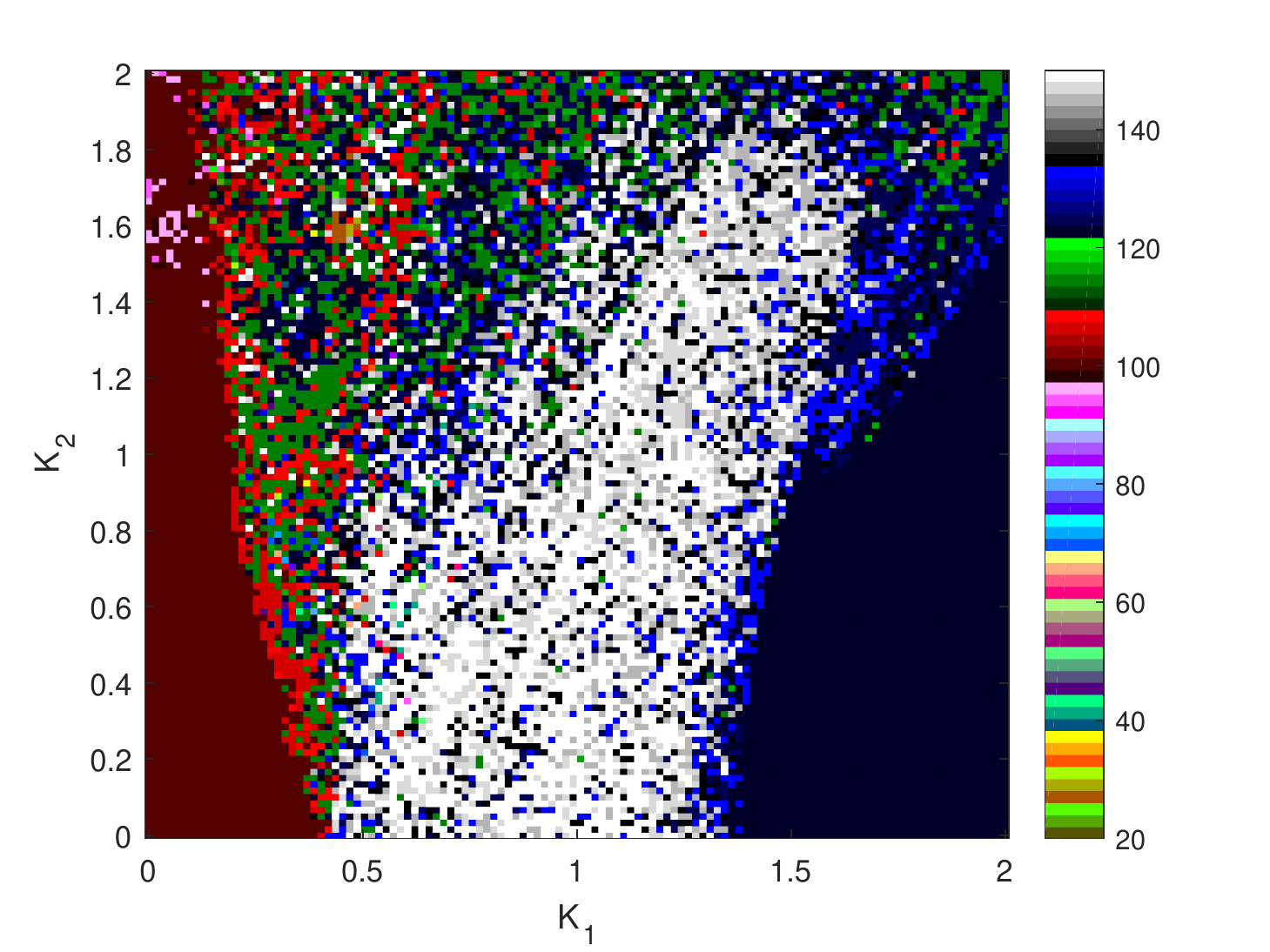}

(c)\hspace{7cm}(d)
\end{center}
\caption{Scan of (a,c) LLEs and (b,d) mean periods for (a,b) van der Pol equation (\ref{e:SCW}) and (c,d) van der Pol Duffing equation (\ref{e:vdPD}) with QP2 forcing (\ref{e:F1forcing}). The forcing amplitudes $k_1$ and $k_2$ of the QP2 forcing are varied on the two axes, while the unforced period is 100~kyr in both cases ($\tau=1$).}
\label{f:vdpdLEQPscan}
\end{figure}

Similar conclusions hold for the two oscillators subject to QP2 forcing (\ref{e:F1forcing}). Recall Figure~\ref{f:forced_vdpvdpd}(c): this shows a typical response of the van der Pol oscillator (\ref{e:SCW}) to two-frequency forcing, while Figure~\ref{f:forced_vdpvdpd}(d) shows a typical response of the van der Pol-Duffing oscillator (\ref{e:vdPD}) to the same forcing. One can numerically verify that the LLE for (c) is negative while that for (d) is positive and chaotic. This is illustrated in Figure~\ref{f:vdpdLEQPscan} which shows (a,c) LLEs and (b,d) mean periods as a function of $k_1$ and $k_2$ for QP2 forcing (\ref{e:F1forcing}) of (a,b) the van der Pol (\ref{e:SCW}) and (c,d) van der Pol-Duffing (\ref{e:vdPD}) systems. Observe the presence of significant regions of chaotic attracting responses for the van der Pol-Duffing system (Figure~\ref{f:vdpdLEQPscan}(c)).

\section{Physics-based conceptual model responses to forcing}
\label{sec:threedim}

We consider two different physics-based models of Pleistocene ice-age oscillations. The first of these is the Saltzmann and Maasch model (SM91) \citep{Saltzman:1991jl} which is a three-variable nonlinear model where all three dynamic variables have similar time scales. The second is the Palliard and Parrenin model (PP04) \citep{Paillard:2004dn} which is a relaxation oscillator (with three dynamic variables) where the switch between glacial and interglacial state is induced by a threshold in Antarctic bottom water formation efficiency. We will see that both SM91 and PP04 models show significant areas in parameter space where a chaotic response to periodic or quasiperiodic forcing (either QP2 or QPn) is possible.

\subsection{The Saltzman and Maasch 1991 model}

The models of Saltzman et al. \citep{Maasch:1990ul,Saltzman:2002tl,Saltzman:1988tv,Saltzman:1990uy,Saltzman:1991jl,Saltzman:1993iq} established the idea that the (late Pleistocene) glacial cycles appear as a limit cycle in the unforced system then synchronised in some way to the orbital forcing. The dynamics (and consequently the specific form of the limit cycle and the bifurcation leading to the limit cycle) varies across the different models. In most of them, it is assumed that the background climate slowly varies throughout the Pliocene-Pleistocene (`tectonically driven' decline in atmospheric CO$_2$-concentration) and the model is formulated as an anomaly model to a (slowly evolving) background climate. \cite{Crucifix2012a} analysed the bifurcation structure of two of these models \citep{Saltzman:1990uy,Saltzman:1991jl} for the full (non-anomaly) equation with respect to one parameter, the slowly drifting `tectonic' CO$_2$ decline $F_\mu(t)$. In these models, much of the interesting dynamics depends on the specific form of the CO$_2$ equation, which at the same time is the most problematic to interpret physically. 

The model SM91 we use here is from \cite{Saltzman:1991jl} and includes one specific representation of the carbon cycle dynamics. The model couples the dynamics of the ice volume $I$, CO$_2$ concentration $\mu$ and deep-ocean temperature $\theta$ as in \cite{Crucifix2012a}:
\ba
\tau\frac{d}{dt} I &=& \alpha_1 - \alpha_2 c\mu - \alpha_3 I - k_{\theta}\alpha_2\theta - k_R\alpha_2 \Lambda(t),\nonumber\\
\tau\frac{d}{dt} \mu &=& \beta_1 -\beta_2\mu + \beta_3\mu^2 -\beta_4 \mu^3 - \beta_5\theta + F_\mu,\nonumber\\
\tau\frac{d}{dt} \theta &=& \gamma_1 -\gamma_2 I - \gamma_3 \theta + F_{\theta}.
\label{e:SM91full}
\ea
The insolation forcing is $\Lambda(t)$ and other parameters are as in Table~\ref{t:SM91}: as in \cite{Crucifix2012a}, we consider the original variables instead of the anomaly variables in that the latter constrained the possible bifurcations and attractors of the system. In \cite{Crucifix2012a}, this is analysed with respect to variations in the parameter $F_{\mu}$, while the other forcings $\Lambda(t)$ and $F_{\theta}$ are set to zero.  We use here fixed values of the `tectonic forcing' $F_{\mu}=F_{\theta}=0$ (see \cite{Crucifix2012a}) and periodic, QP2 or QPn forcing $\Lambda(t)$. We fix $k_R=0.4$ to give comparability with other models in terms of response to forcing.

\begin{figure}
	\begin{center}
		\includegraphics[width=8cm]{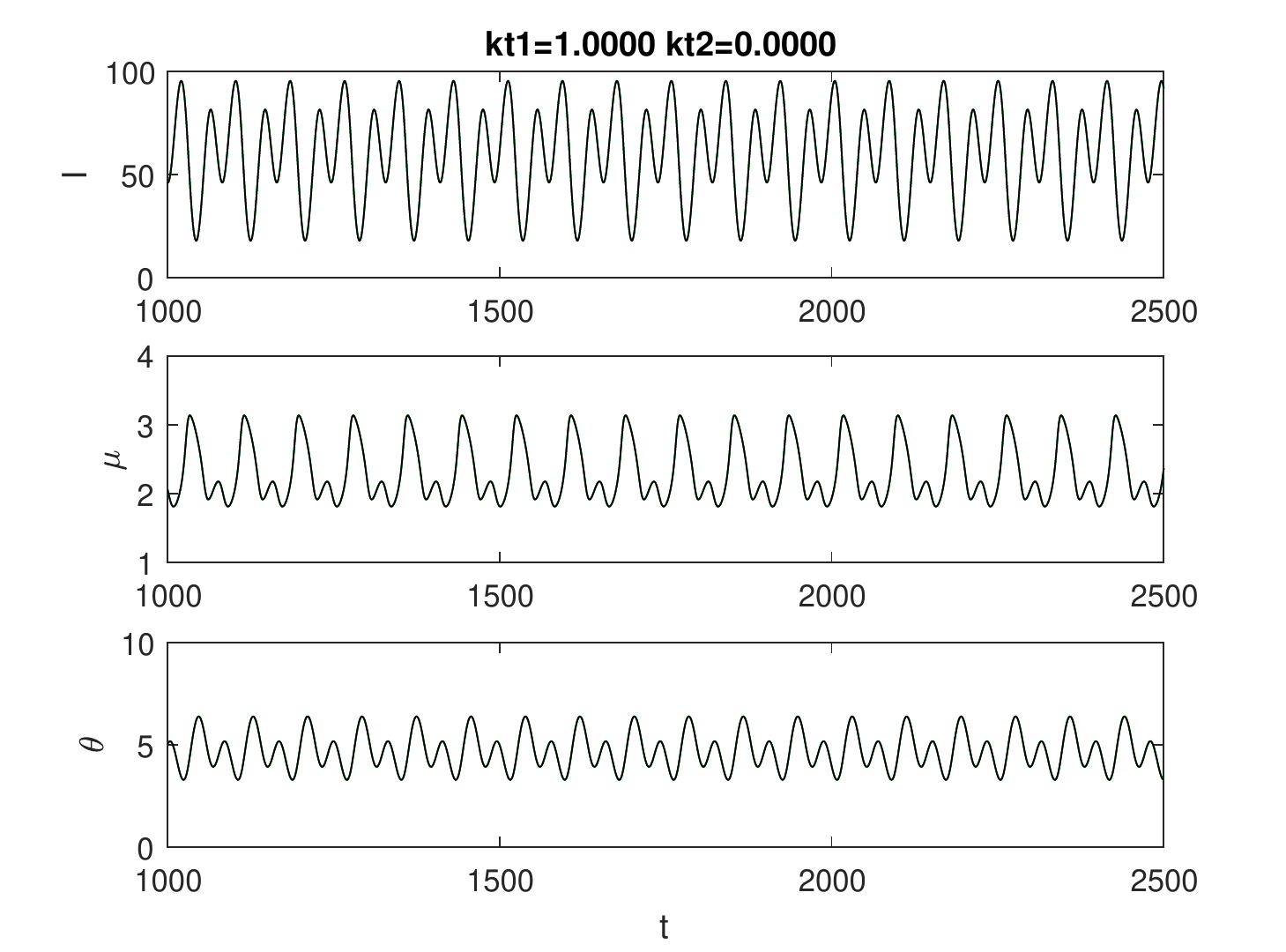}~
		\includegraphics[width=8cm]{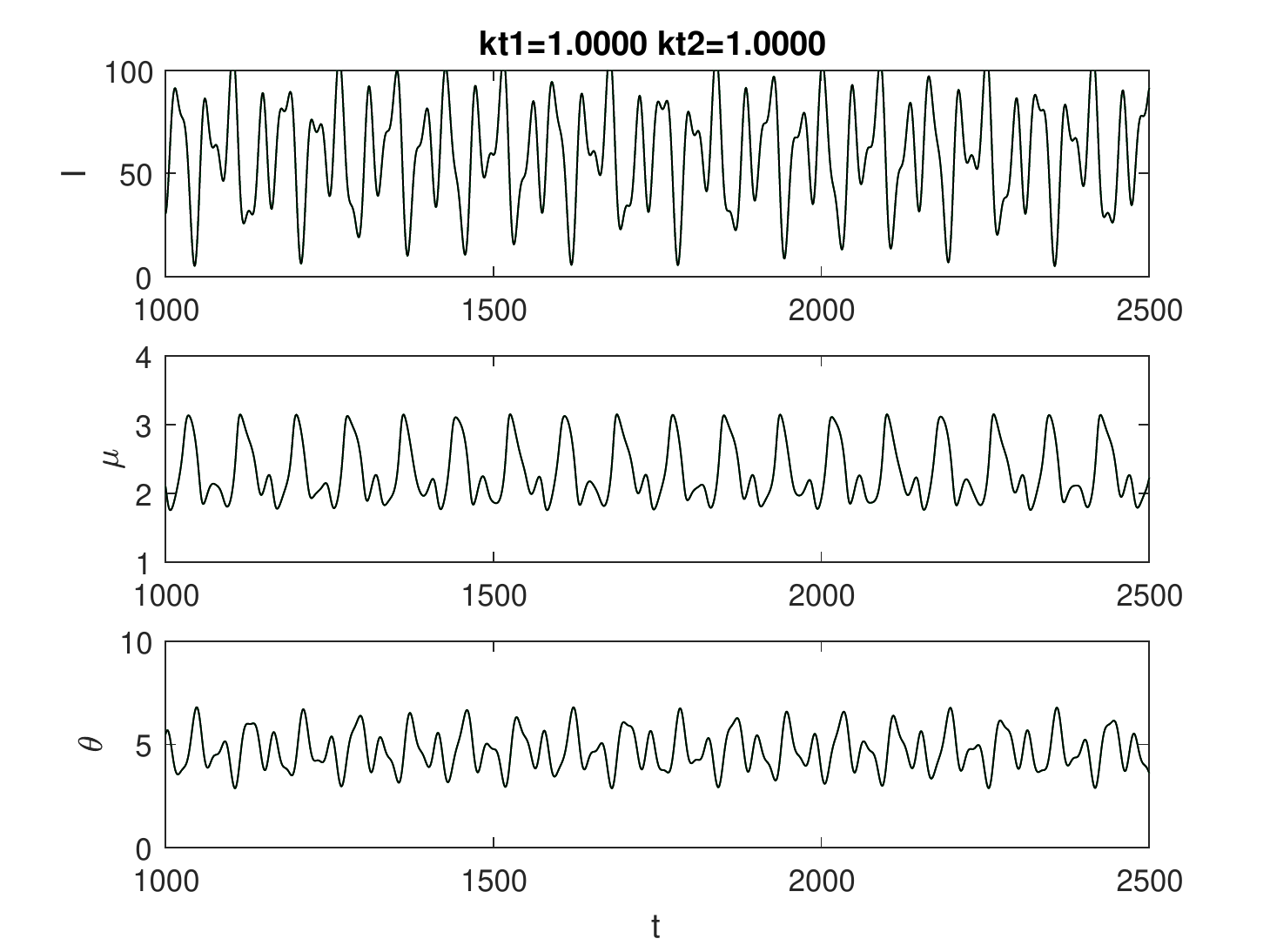}~
		
		(a)\hspace{8cm}(b)

		\includegraphics[width=8cm]{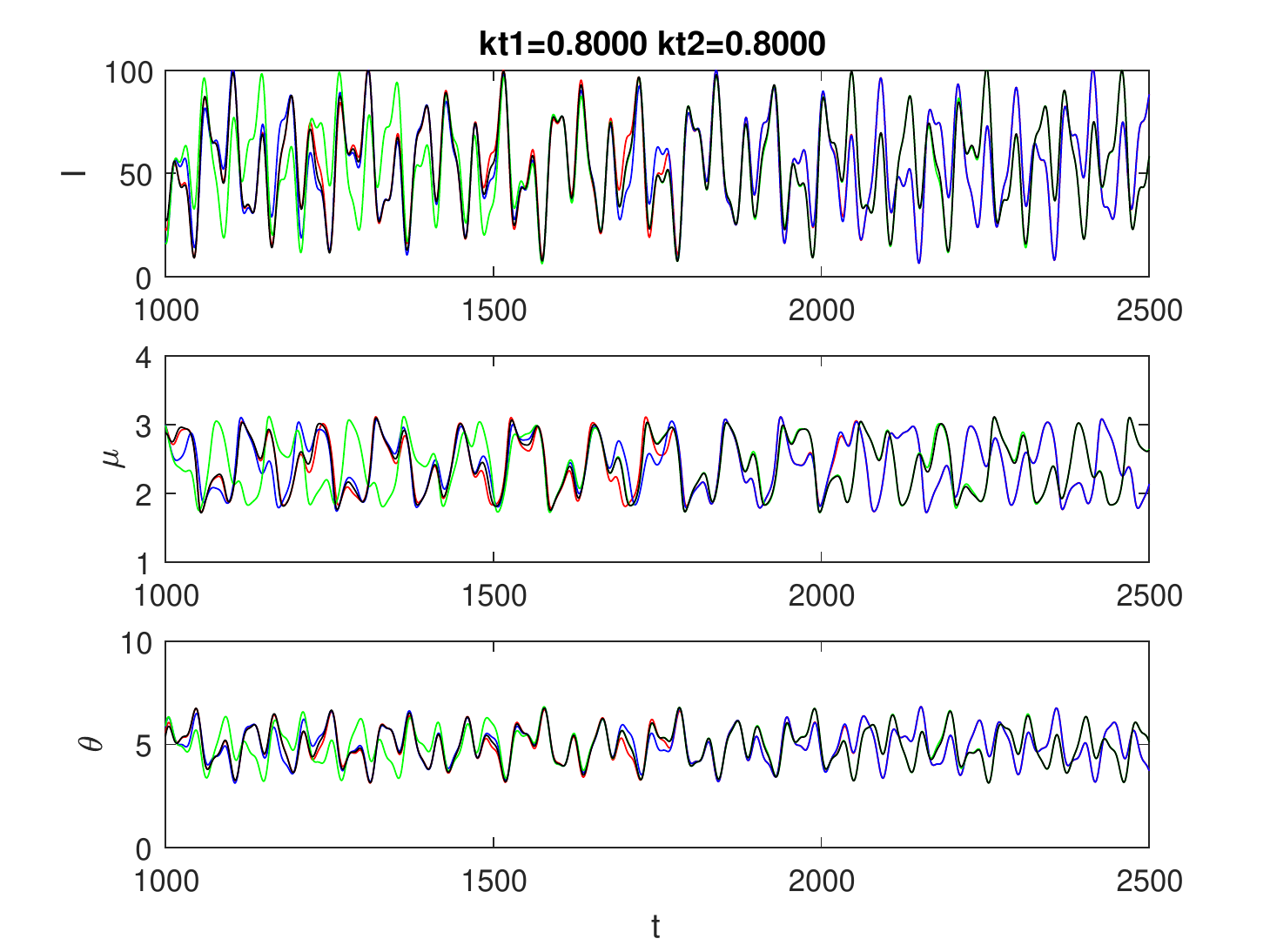}~
		\includegraphics[width=8cm]{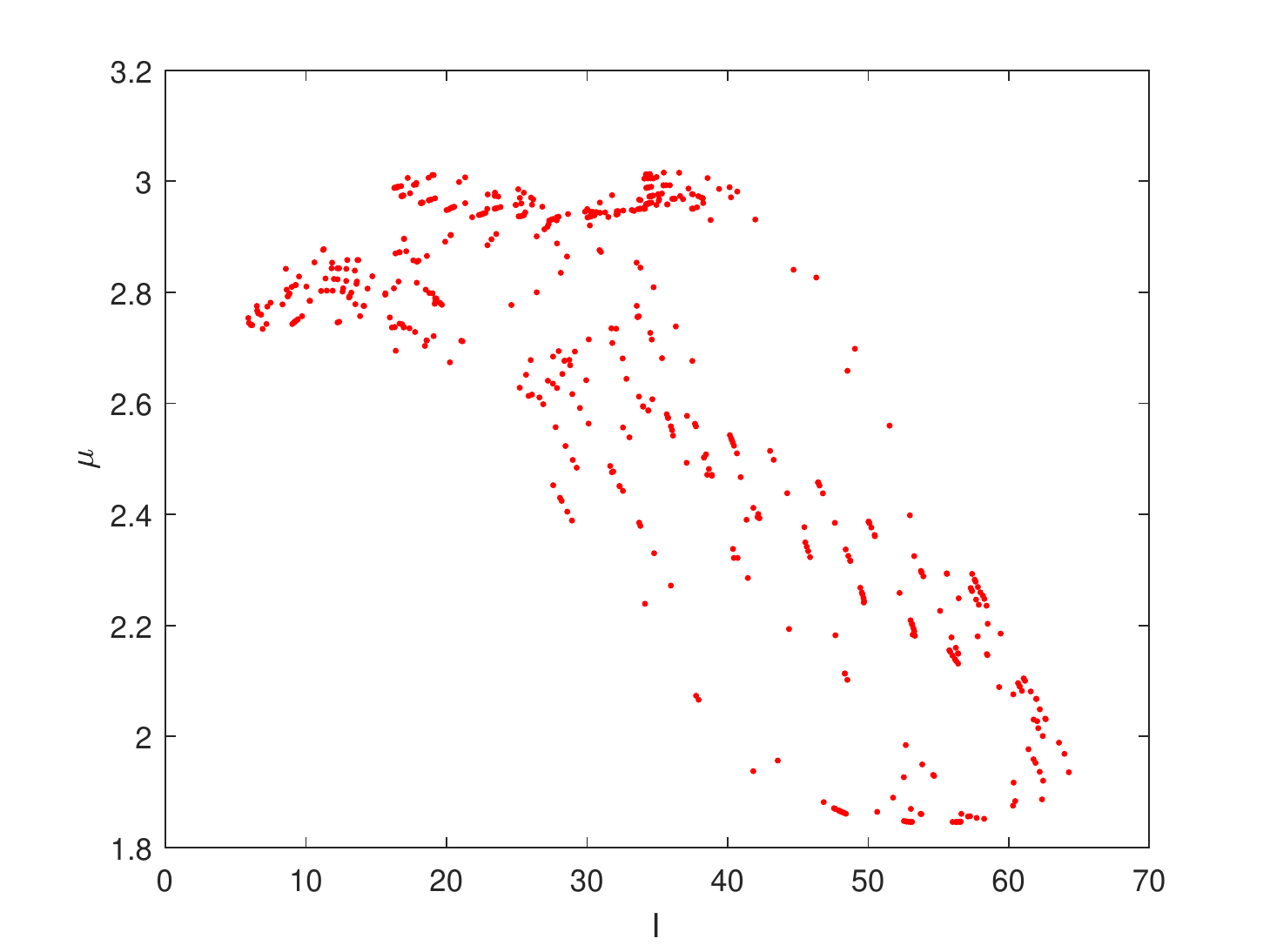}~
		
		(c)\hspace{8cm}(d)

	\end{center}
	\caption{Examples of time series for four randomly chosen initial conditions (shown in different colours) for the SM91 model (\ref{e:SM91full}) with parameters as in Table~\ref{t:SM91}. Panel (a) shows responses to periodic forcing (\ref{e:F0forcing}) with $k_1=1$ and (b) shows response to QP2 forcing (\ref{e:F1forcing}) with $k_1=k_2=1$. In both cases there is synchronization to a single trajectory. Panel (c) shows QP2 forcing with $k_1=0.8$, $k_2=0.8$ and a response consistent with chaos. The Poincar\'{e} section (d) corresponds to one of the trajectories in (c) sampled at zero phase of the obliquity forcing.}
	\label{f:SMTS}
\end{figure}

Figure~\ref{f:SMTS}(a) illustrates time series of a typical response of this system (\ref{e:SM91full}) to periodic forcing (\ref{e:F0forcing})  while (b) shows a corresponding response to QP2 forcing (\ref{e:F1forcing}). The evolution of four initial conditions are shown on the same axes - note that in both of these cases there is synchronization to the same trajectory, typical of an attractor with a negative LLE.
For $k_1=0$ (precession only) there is also locking present for larger $k_2$ (not shown). Note the presence of a region of chaotic locking to approximately 100~kyr with positive LLEs: Figure~\ref{f:SMTS}(c) shows an example of where the four time series repeatedly and intermittently synchronize for periods of time while (d) shows a projected Poincar\'{e} section sampled when the phase of obliquity is zero modulo $2\pi$ for a longer single trajectory from (c).

\begin{figure}
	\begin{center}
		\includegraphics[width=0.48\textwidth]{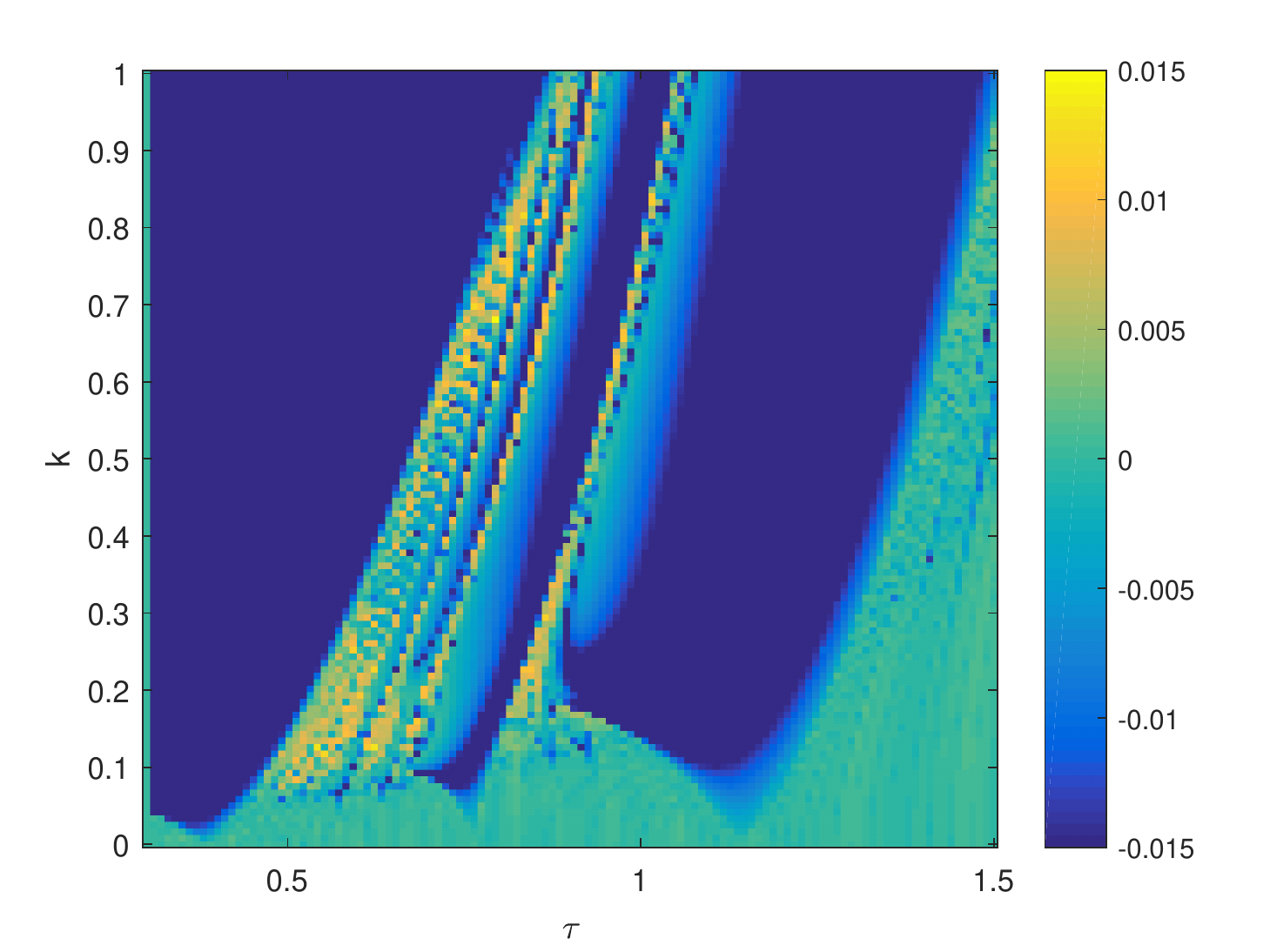}~
		\includegraphics[width=0.48\textwidth]{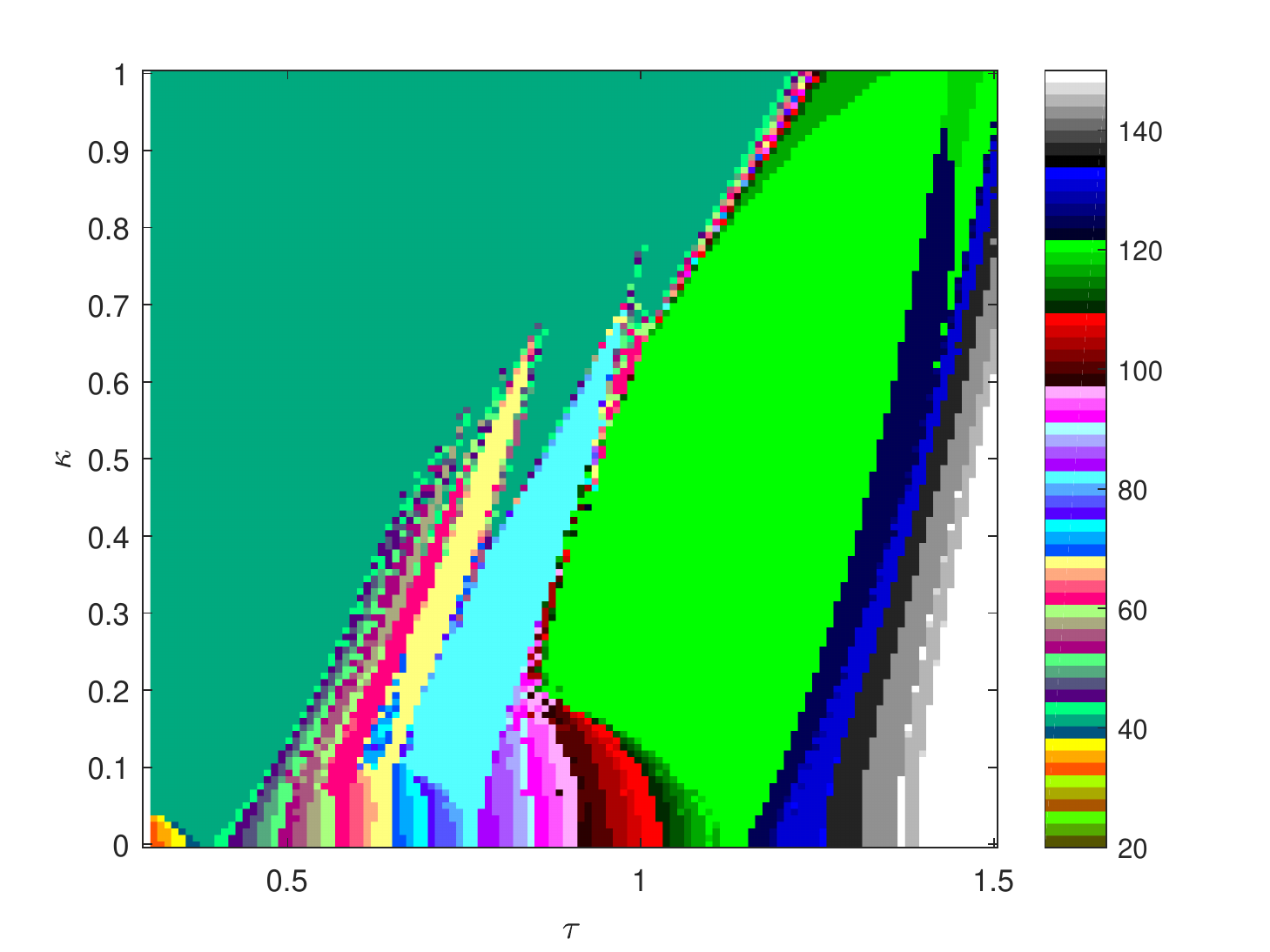}
		
		(a)\hspace{7cm}(b)
		\includegraphics[width=0.48\textwidth]{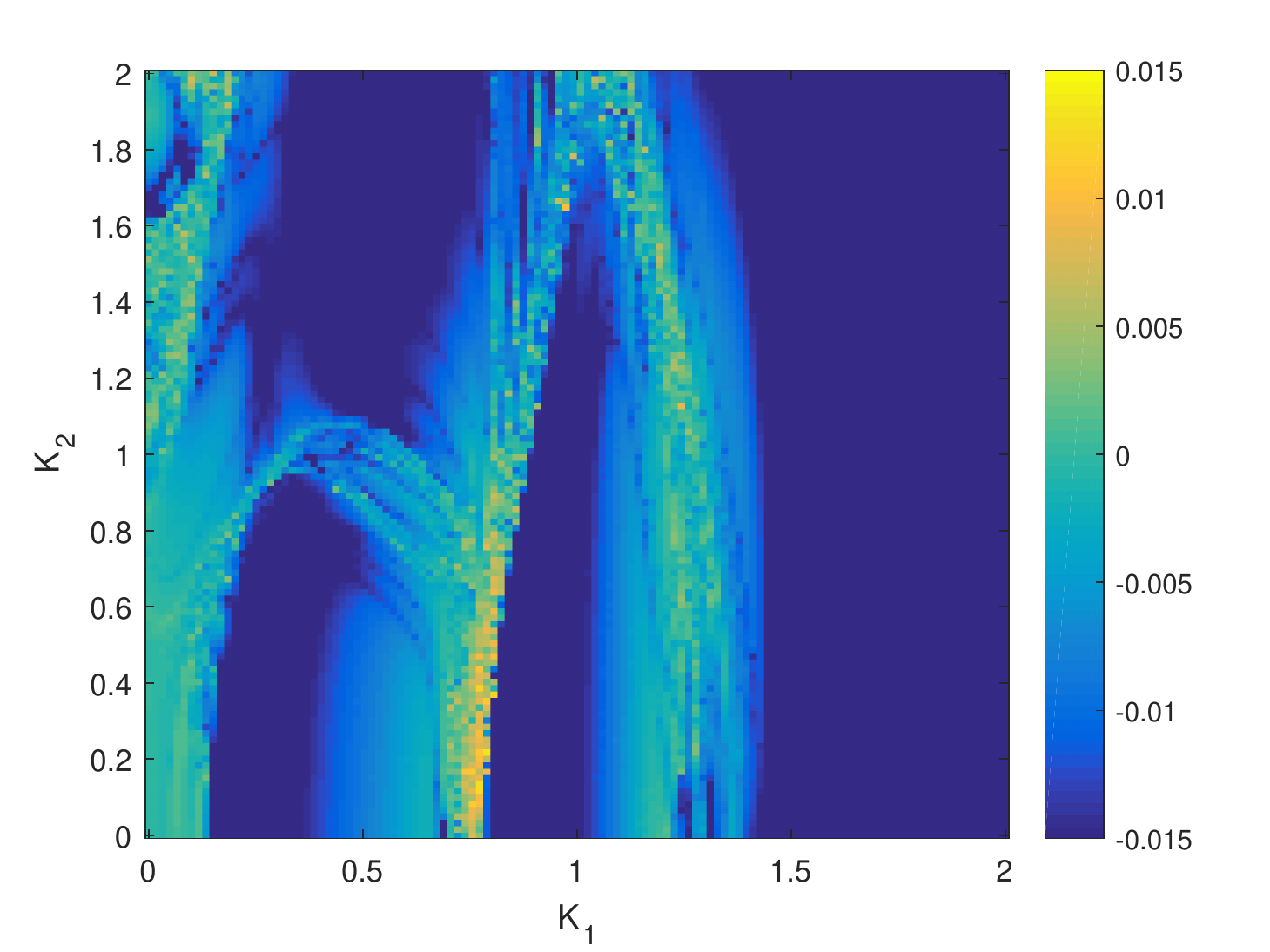}~
		\includegraphics[width=0.48\textwidth]{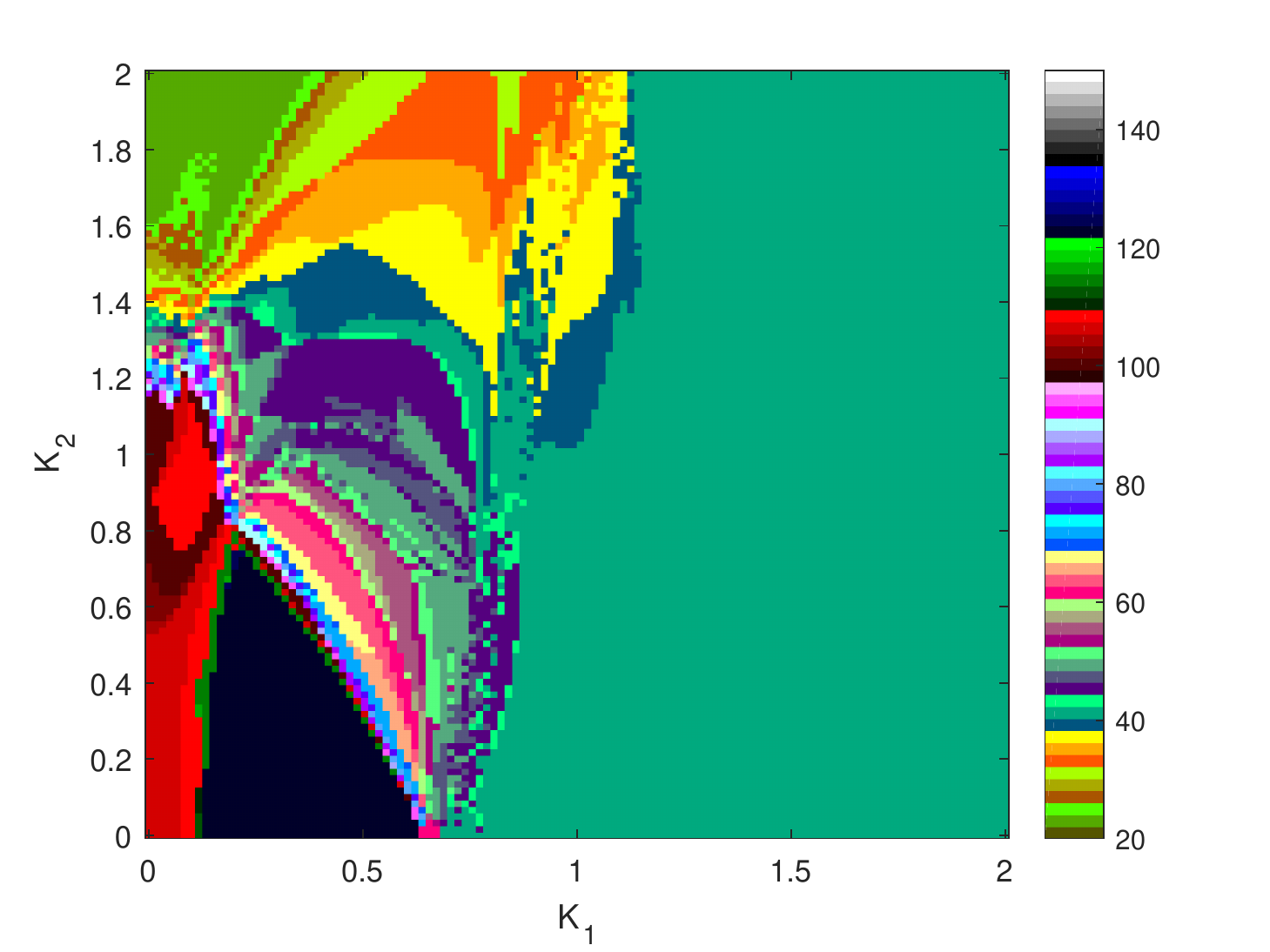}
		
		(c)\hspace{7cm}(d)

      \end{center}
	\caption{Scan (a,c) of LLEs and (b,d) mean periods for SM91 Saltzmann/Maasch model (\ref{e:SM91full}). Panels (a,b) show the responses for periodic forcing (\ref{e:F0forcing}), plotting forcing amplitude $\kappa=k_1$ against $\tau$. Observe the Arnold tongue structure for $k_1$ small and the presence of chaotic responses for $k_1$ larger. Panels (c,d) show the responses for QP2 forcing (\ref{e:F1forcing}), plotting obliquity $k_1$ and the precessional $k_2$ forcing amplitude. Observe the presence of regions of approximately 100~kyr responses, and regions where there are positive LLEs.}
	\label{f:SMLEscanQP}
\end{figure}

Figure~\ref{f:SMLEscanQP}(a,c) illustrates LLEs and (b,d) mean periods for a scan with periodic forcing (\ref{e:F0forcing}) for this model. Panels (a,b) show scans for varying time scaling $\tau$ (which scales the unforced oscillation period) and forcing amplitude $\kappa=k_1$ while panels (c,d) show scans through amplitudes of the components for QP2 forcing (\ref{e:F1forcing}). There are clearly regions of chaotic behaviour in parameter space, mostly concentrated at larger values of the precessional component amplitude $k_2=1$ and $k_1$ small. For $k_2=0$ (obliquity only) observe that there is a large region of 1:1 locking to the obliquity for larger $k_1$, and a region of 3:1 locking for smaller $k_1$.

\begin{table}[b!]
\begin{center}
	\begin{tabular}{|llll|}
		\hline
		 & Value & Units & Physical interpretation \\
		\hline
		$\alpha_1$ & $1.673915\times10^{16}$ & kg yr$^{-1}$  & Constant ice growth rate. \\
		$\alpha_2$ & $9.52381\times10^{15}$ & kg yr$^{-1}$ & Effect of CO$_2$ and $\theta$ on ice dynamics.\\
		$\alpha_3$ & $10^{-4}$ & yr$^{-1}$ & Inverse linear response time ice.\\
		$\beta_1$  & $0.5118377$ & ppm yr$^{-1}$& CO$_2$ coefficients.\\
		$\beta_2$ & $ 6.258680\times 10^{-3}$ & yr$^{-1}$ & CO$_2$ coefficients.\\
		$\beta_3$ & $2.639456\times 10^{-5}$ & (ppm yr)$^{-1}$ & CO$_2$ coefficients.\\
		$\beta_4$ & $3.628118\times 10^{-8}$ & (ppm$^2$ yr)$^{-1}$   & CO$_2$ coefficients.\\
		$\beta_5$ & $5.833333\times 10^{-3}$ & ppm $(\degree$C yr$)^{-1}$ & CO$_2$ coefficients.\\
		$\gamma_1$ & $1.85125\times 10^{-3}$ & $\degree$C yr$^{-1}$ & Rate of growth of $\theta$.\\
		$\gamma_2$ & $1.125\times 10^{-23}$ & $\degree$C(kg yr)$^{-1}$ & Effect of ice on $\theta$.\\
		$\gamma_3$ & $2.5\times 10^{-4}$ & yr$^{-1}$ & Inverse linear response time $\theta$.\\
		$c$ & $4\times 10^{-3}$ & ppm$^{-1}$ & \\
		$\kappa_\theta$ & $4.4444444\times 10^{-2}$ & ($\degree$C)$^{-1}$ &\\
		$\kappa_R$ &  $0.4$ & & Insolation normalization\\
		$\tau$ & 1 & & Time scaling \\
		\hline
	\end{tabular}
	\end{center}
	\caption{Parameters and their interpretation for the model \cite{Saltzman:1991jl} (SM91). Where possible, we use the parameter values of \cite{Crucifix2012a} in the analysis of the full model. }
	\label{t:SM91}
\end{table}

\subsection{The Paillard-Parrenin 2004 model}

The model of \cite{Paillard:2004dn} (PP04) gives an explicit qualitative physical explanation for switches between glacial and interglacial regimes. The key assumption here is that deglaciations are induced by changes in atmospheric CO$_2$. In this case, the (slow) transition from interglacial to glacial can be explained by Milankovitch theory, where the summer insolation in northern high latitudes play a major role. The faster glacial to interglacial transition (climate escapes from a deep glacial) is caused by the release of CO$_2$ from the deep ocean into the atmosphere.  In this model, this is achieved by an oceanic switch, characterised by a salty bottom water formation efficiency parameter $F$. During cold periods, the Southern Ocean bottom water formation is strong, the deep ocean is strongly stratified with cold and salty Antarctic bottom water (AABW) and can store a lot of carbon. However, a few thousand years after a glacial maximum, the Antarctic ice sheet reaches its maximum extent making salty AABW formation through brine rejection on the continental shelf more difficult, the Southern Ocean stratification weakens while at the same time atmospheric CO$_2$ rises rapidly. 

The model involves three variables, the global ice volume $V$ forced by CO$_2$ and Northern Hemisphere summer insolation, the Antarctic ice sheet extent $A$ driven by sea level changes (via $V$) and the atmospheric CO$_2$ concentration $C$ connected to the deep water state (glacial or interglacial):
\ba
\tau \frac{d}{dt} V &=& \left(-aC-bI_{65N}+c-V\right)/\tau_{V},\nonumber\\
\tau \frac{d}{dt} A &=& \left(V-A\right)/\tau_{A},\nonumber\\
\tau \frac{d}{dt} C &=& \left(d I_{65N}-e V + f \mathcal{H}(-F)+g-C\right)/\tau_C.
\label{e:PP04}
\ea
The oceanic (salty bottom water) switch parameter $F$ is given by:
\be
F = h V - i A +j.
\label{e:PP04switch}
\ee
The forcing $I_{65N}=\Lambda(t)$ and the parameter values are given in Table~\ref{t:PP04}. We approximate the Heaviside function in (\ref{e:PP04}) using $\mathcal{H}(x)=(1+\tanh(Kx))/2$ with $K=100$. It should be noted that the only nonlinearity in this system is the Heaviside function (or its tanh approximation) which provides a state-dependent switch between two linear modes: (i) an ice-accumulation mode where the system relaxes towards a low CO$_2$, high ice volume state and (ii) an ice-ablation mode where the systems relaxes towards a high CO$_2$, low ice volume state.

\begin{table}[h!]
\begin{center}
		\begin{tabular}{|llll|}
			\hline
			Parameter & Value & Units & Physical interpretation\\
			\hline
			$\tau_V$ &  15 & kyr &  Time scale global ice volume\\
			$\tau_A$ & 12 & kyr & Time scale Antarctic ice sheet\\
			$\tau_C$ &  5 & kyr & Time scale carbon cycle\\
			$a$ &  1.3 & & driving $V$ with $C$ \\
			$b$ &  0.5 & & driving $V$ by insolation\\ 
			$c$ &  0.8 & & \\
			$d$ &  0.15 & & driving $C$ with insolation \\
			$e$ &  0.5 & & driving $C$ with $V$ \\
			$f$ &  0.5 & & strength of ocean switch\\
			$g$ &  0.4 & & \\
			$h$ &  0.3 & & salty bottom water efficiency\\
			$i$ &  0.7 & & \\
			$j$ &  0.27 & & \\
			$\tau$ & 1 & & Time scaling \\
			\hline
	\end{tabular}
	\end{center}
	\caption{Chosen values of parameters and their interpretation for the PP04 model (\ref{e:PP04}) of \cite{Paillard:2004dn}: these correspond to the parameter values used in \cite{Crucifix2012a}. As $V$, $A$ and $C$ are dimensionless quantities, most parameters are dimensionless. }
		\label{t:PP04}
\end{table}

\begin{figure}
	\begin{center}
		\includegraphics[width=8cm]{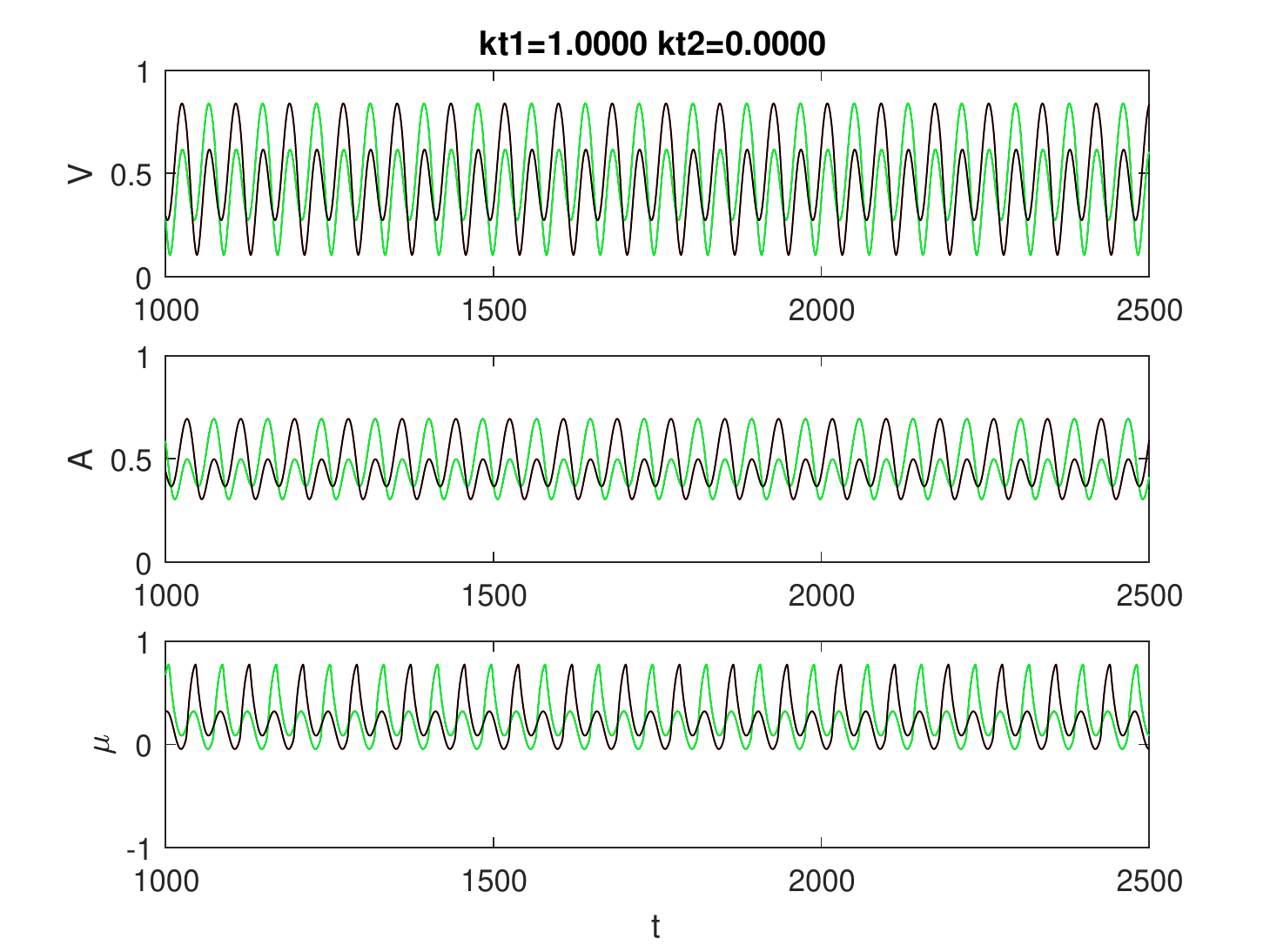}
		\includegraphics[width=8cm]{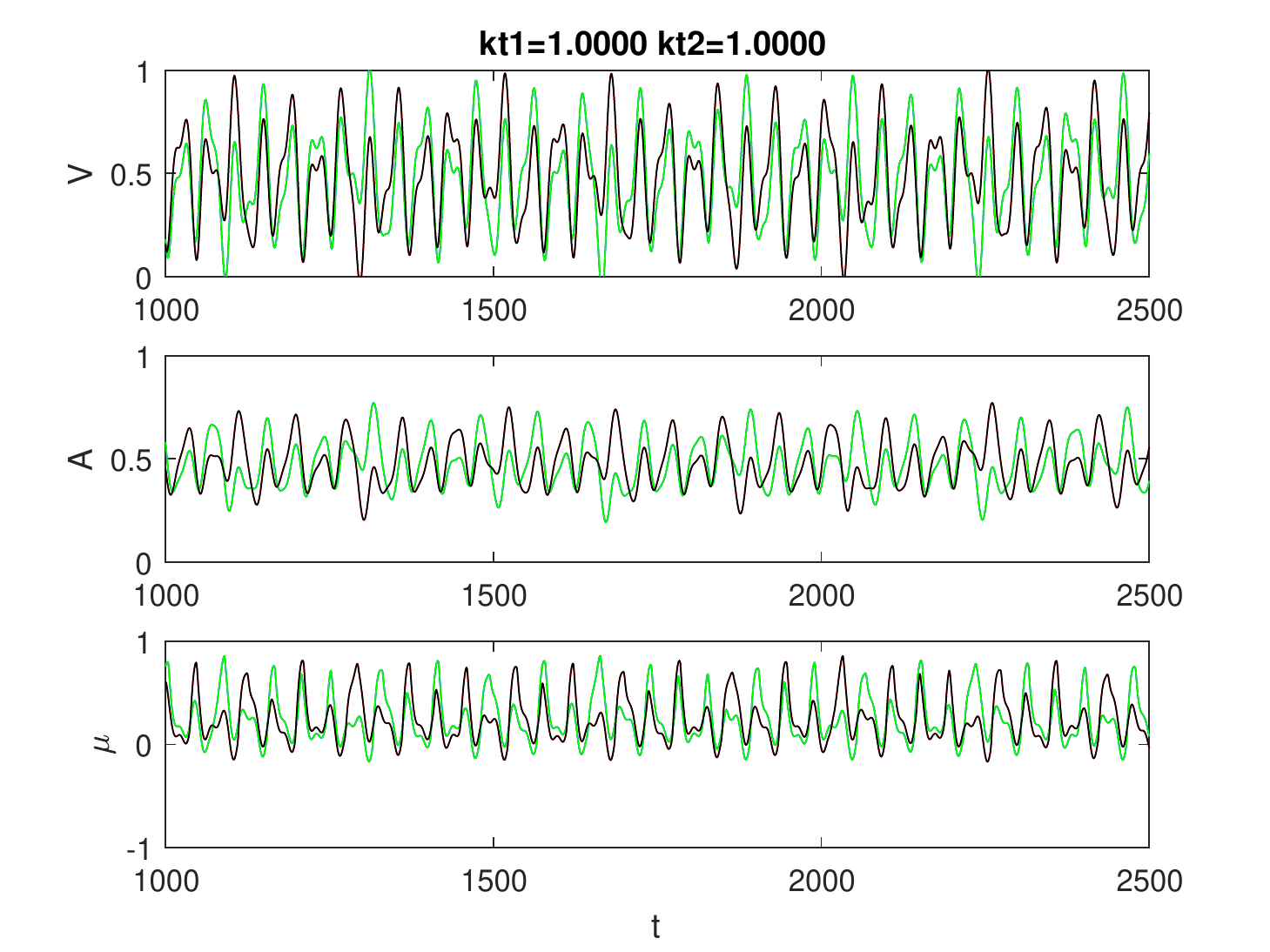}
		
		(a)\hspace{8cm}(b)

		\includegraphics[width=8cm]{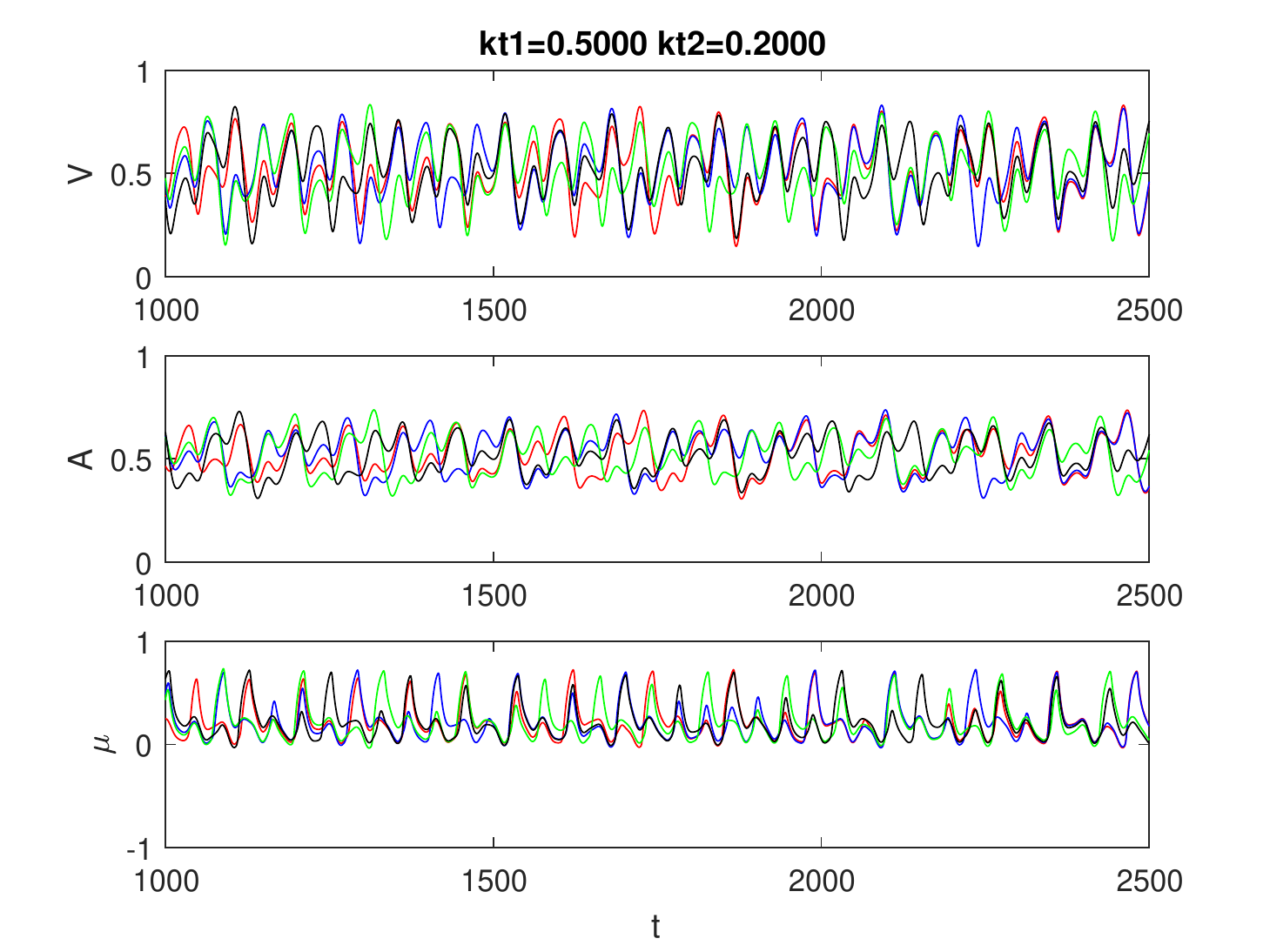}
		\includegraphics[width=8cm]{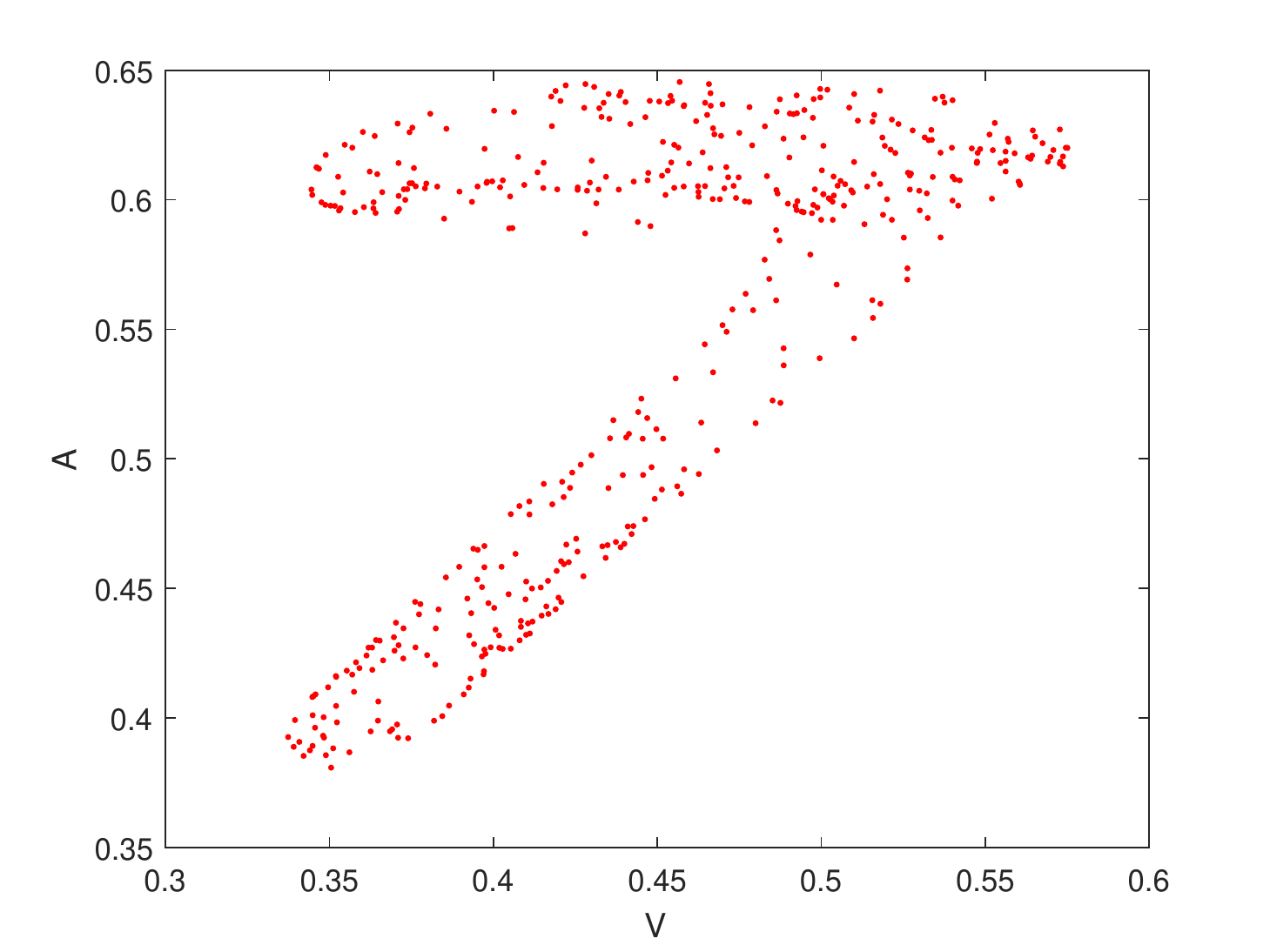}
		
		(c)\hspace{8cm}(d)

	\end{center}
	\caption{Examples of time series for four randomly chosen initial conditions (shown in different colours) for the PP04 (\ref{e:PP04}) with parameters as in Table~\ref{t:PP04}: observe in this case there is no 1:1 locking but rather a subharmonic locking. Panel (a) shows responses to periodic forcing (\ref{e:F0forcing}) with $k_1=1$ and (b) shows response to QP2 forcing (\ref{e:F1forcing}) with $k_1=k_2=1$. Panel (c) shows QP2 forcing with $k_1=0.5$, $k_2=0.2$. Poincar\'{e} section (d) corresponding to one of the trajectories in (c) sampled at zero phase of the obliquity forcing.}
	\label{f:PPTS}
\end{figure}

\begin{figure}
	\begin{center}
		\includegraphics[width=0.48\textwidth]{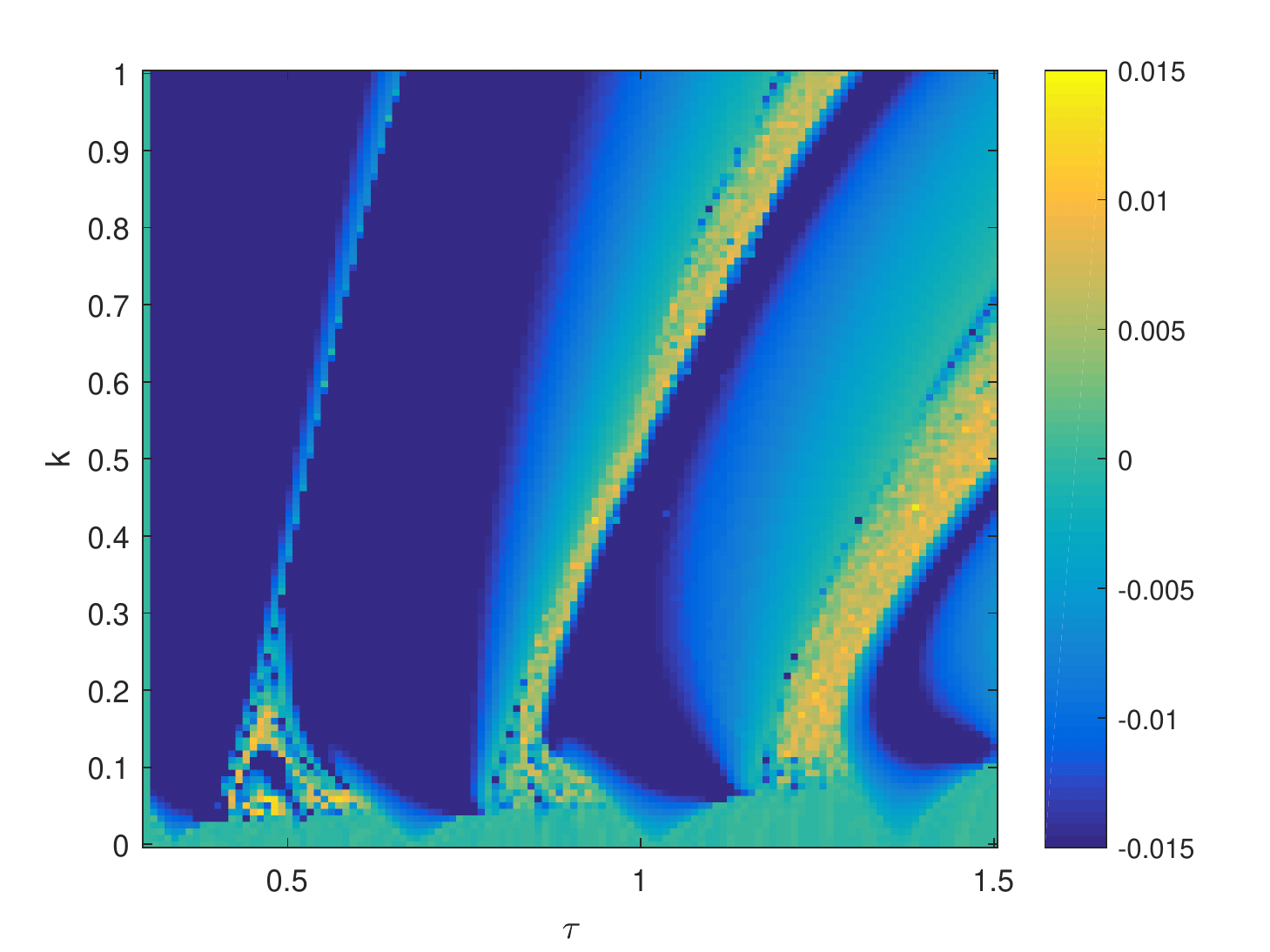}~
\		\includegraphics[width=0.48\textwidth]{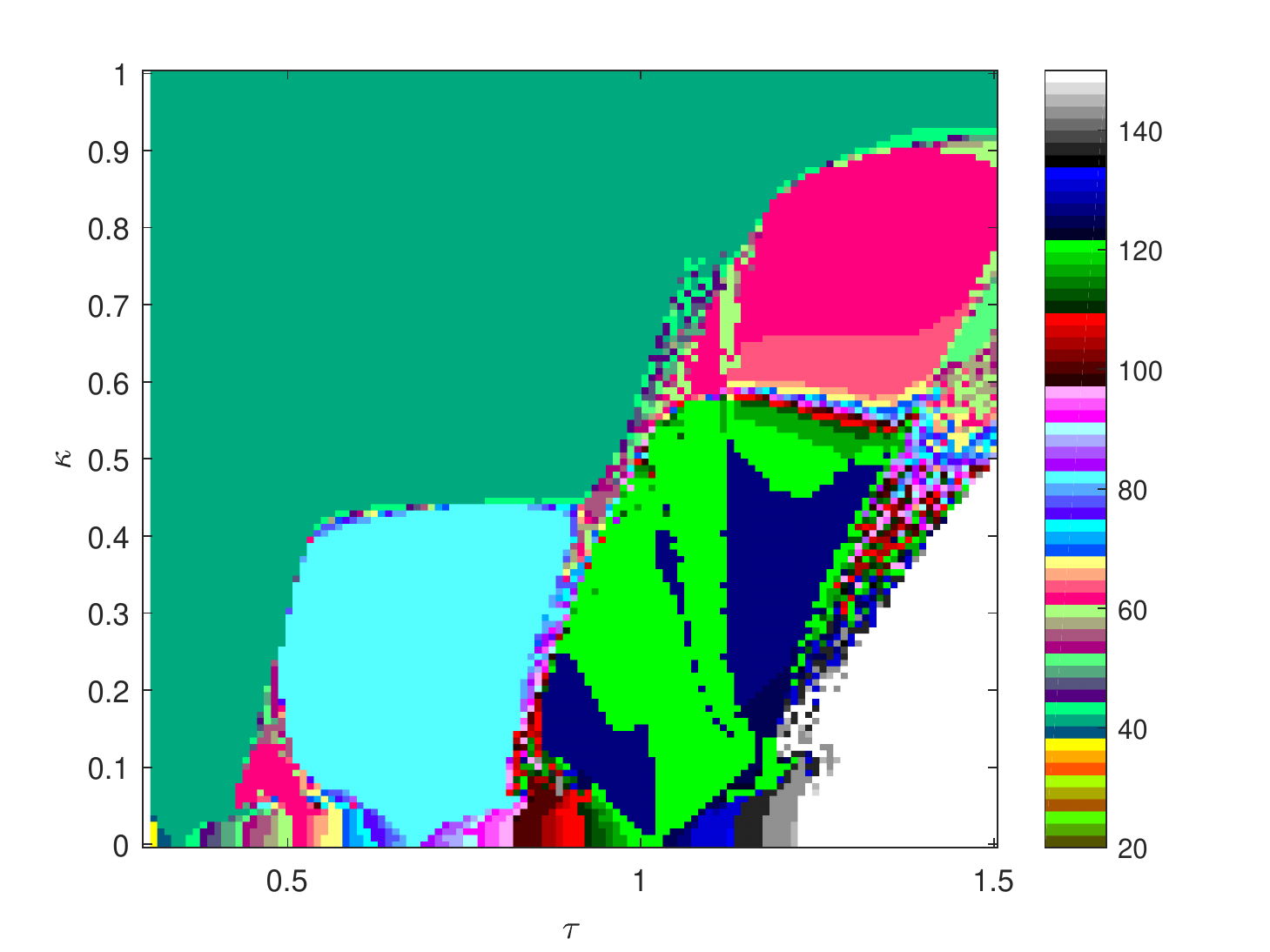}
		
		(a)\hspace{7cm}(b)

		\includegraphics[width=0.48\textwidth]{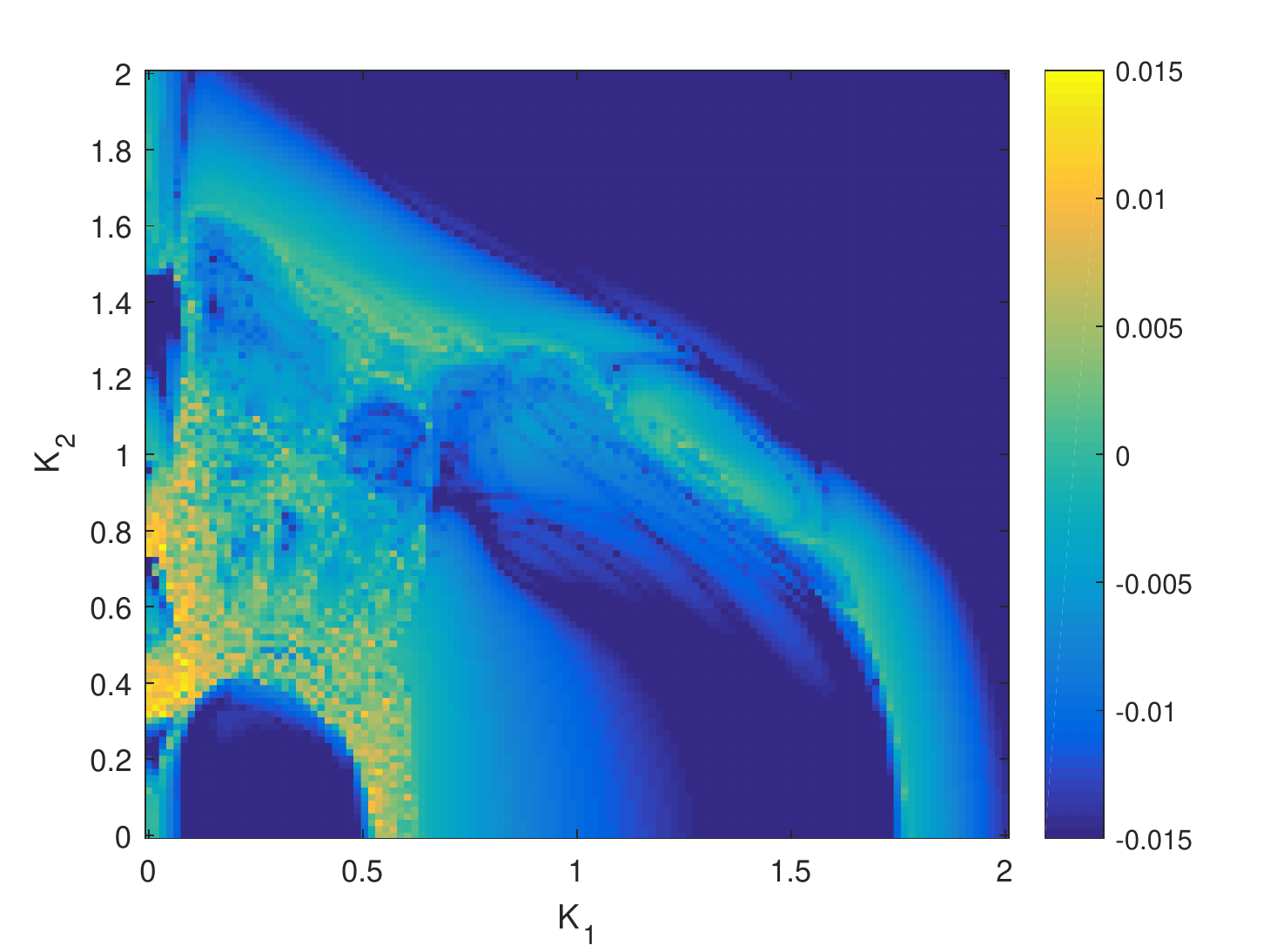}~
		\includegraphics[width=0.48\textwidth]{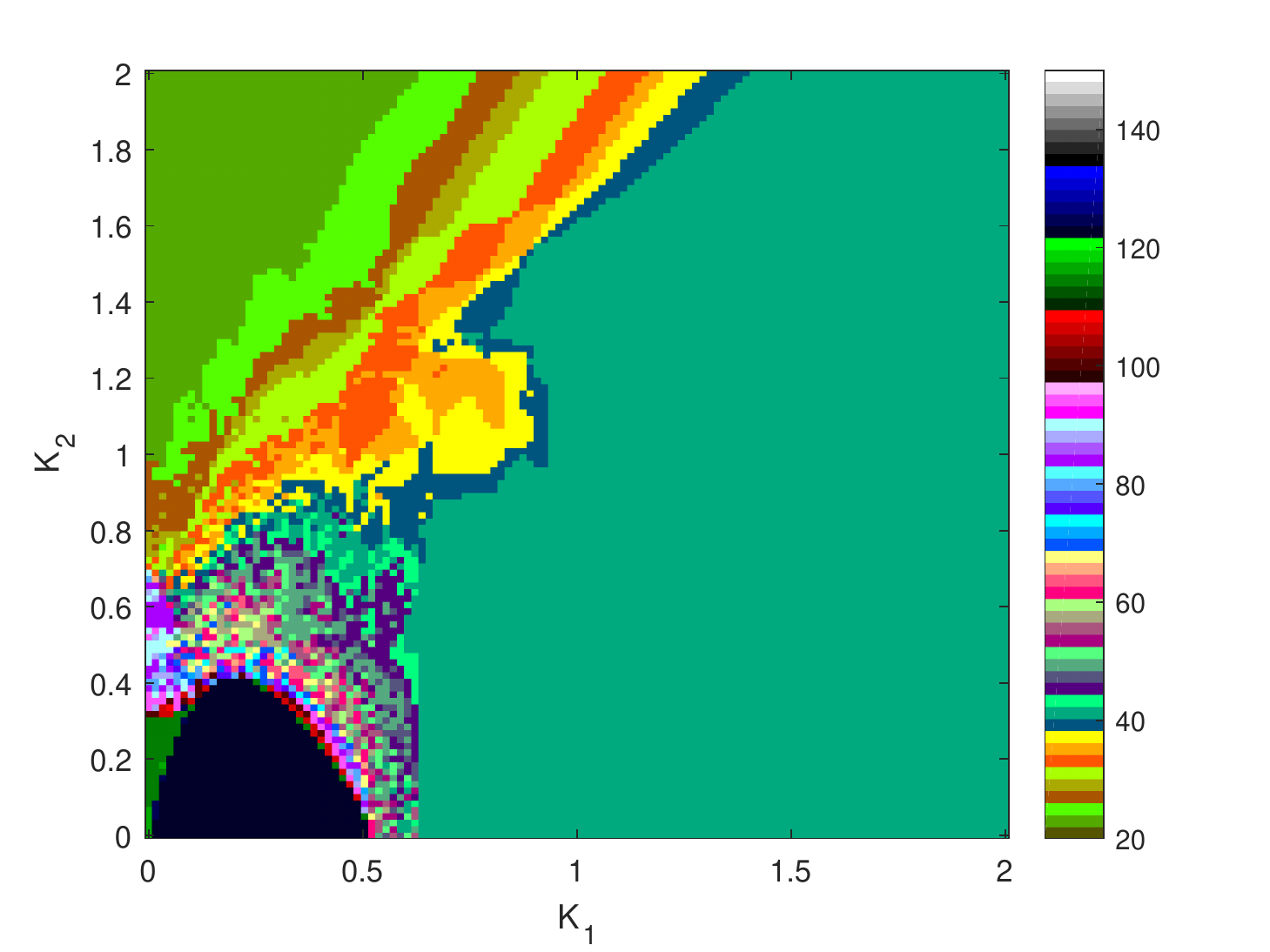}
		
		(c)\hspace{7cm}(d)

	\end{center}
	\caption{Scan of (a,c) LLEs and (b,d) mean periods for the PP04 Palliard/Parrenin model (\ref{e:PP04}). Panels (a,b) show response to periodic forcing (\ref{e:F0forcing}), scanning through $\kappa=k_1$ and $\tau$. Panels (c,d) show response to QP2 forcing (\ref{e:F1forcing}) on varying the amplitude of the forcings $k_1$ and $k_2$. There are regions of approximately 100~kyr mean period and regions where there are positive LLEs.}
	\label{f:PPLEscanQP}
\end{figure}

The unforced system exhibits internal oscillations at a 132~kyr period for the parameters used by \cite{Paillard:2004dn}. Figure~\ref{f:PPTS}(a) illustrates time series of a typical response of this system to periodic forcing - in this case there are several attractors corresponding to mode-locking to different phases. Figure~\ref{f:PPTS}(b) show responses of this system to QP2 forcing (\ref{e:F2forcing}) with $k_1=k_2=1$ while (c) shows the responses for $k_1=0.5$, $k_2=0.2$ and an example with apparent chaotic response (see also Figure~\ref{f:LLEexample} in the Appendix). Computing LLEs for a scan through $k_1$ the amplitude of the obliquity and $k_2$ precessional components, 

Figure~\ref{f:PPLEscanQP}(a,b) shows the response of PP04 (\ref{e:PP04}) to periodic forcing (\ref{e:F0forcing}) on varying the time scaling $\tau$ and amplitude $\kappa=k_1$: (a) shows the LLE (b) the mean period. Similar to the SM91 model Arnold tongues appear for small $\kappa$ in this model and Figure~\ref{f:PPLEscanQP}(c) shows regions of positive LLE and (d) mean period for QP2 forcing (\ref{e:F1forcing}) on varying the amplitudes $k_1$ and $k_2$.

\begin{figure}
	\begin{center}
		\includegraphics[width=0.48\textwidth]{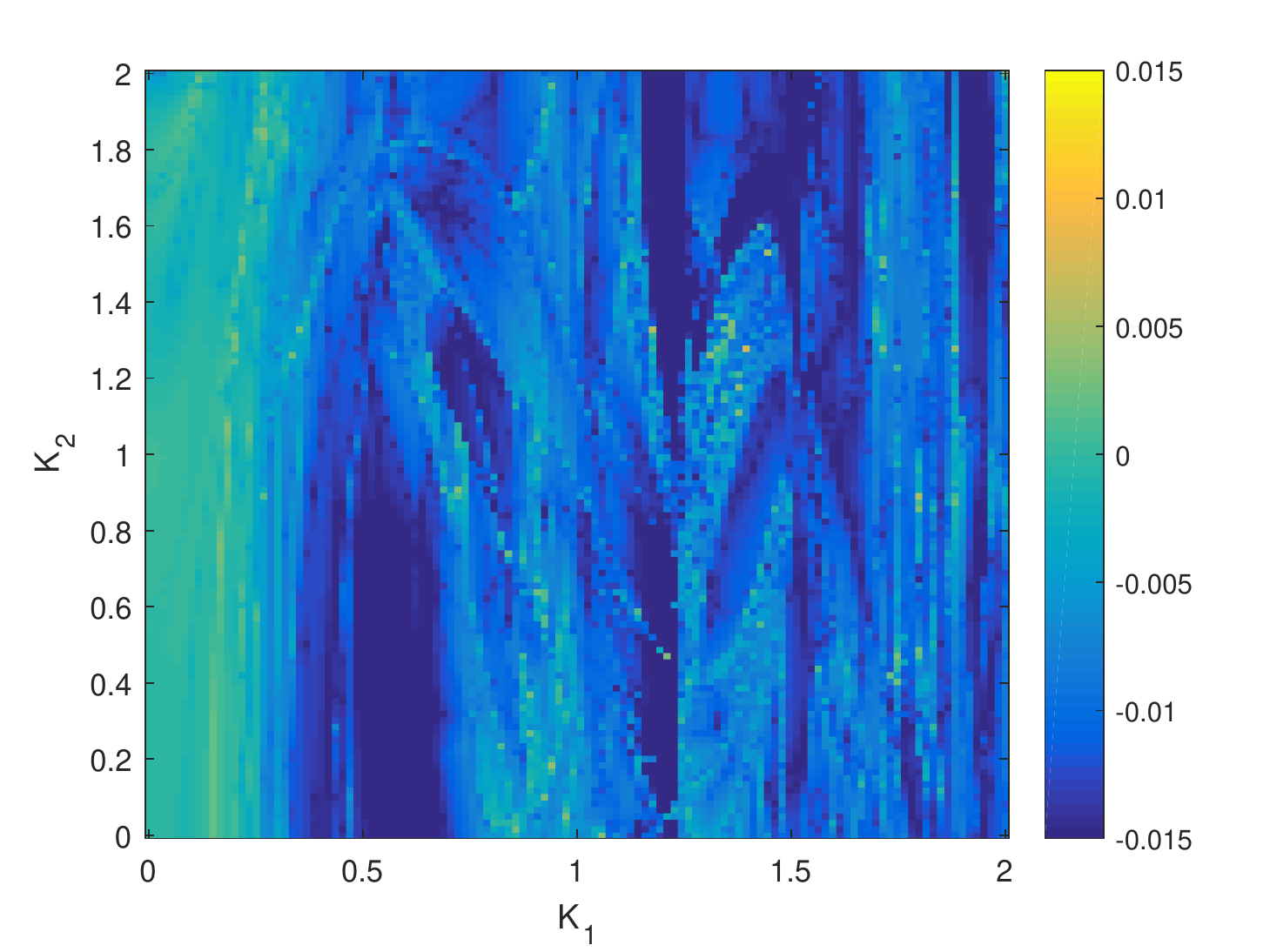}~
		\includegraphics[width=0.48\textwidth]{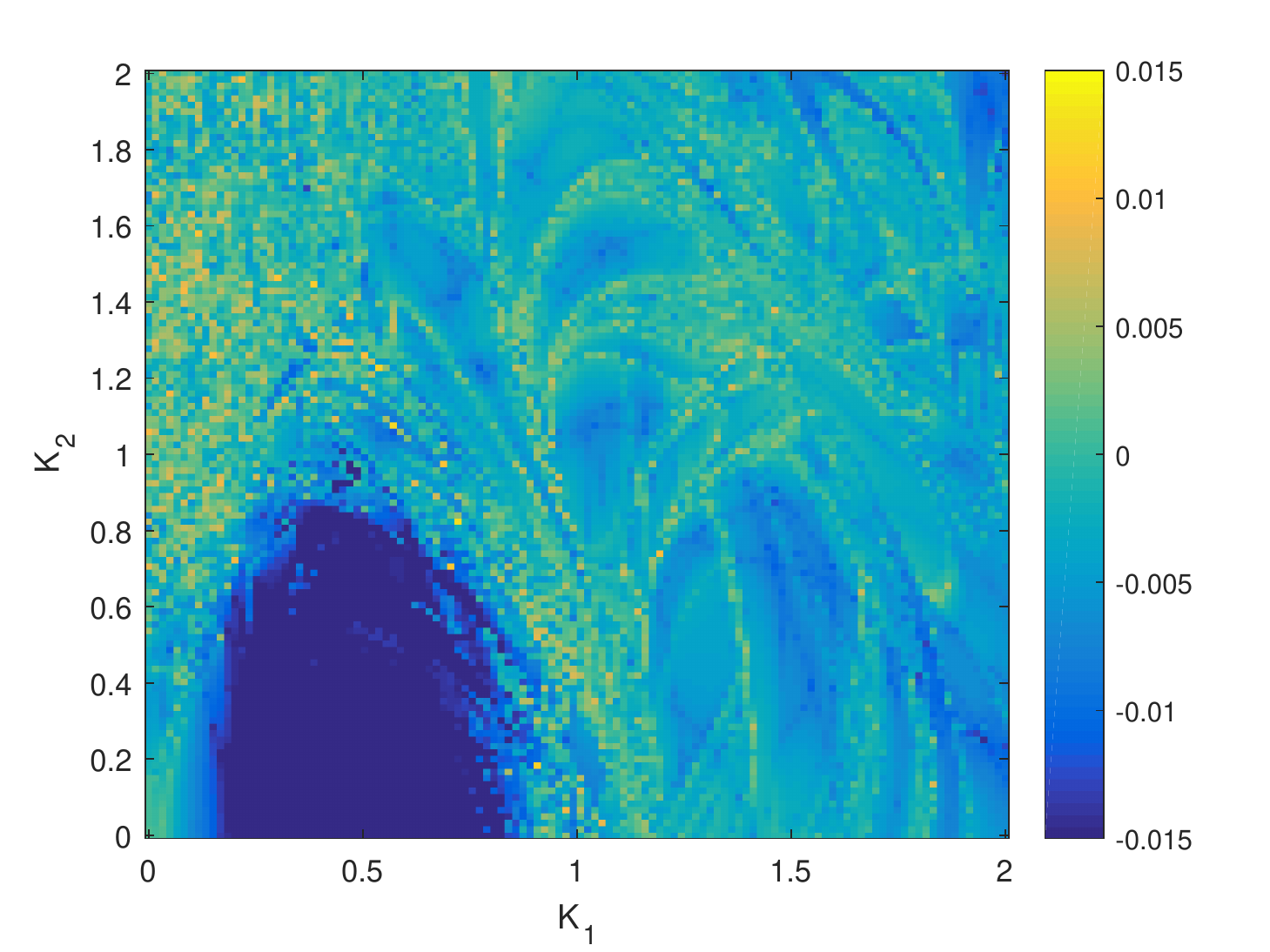}~

		(a)\hspace{7cm}(b)

		
		\end{center}
	\caption{Scans of LLEs for (a) Saltzmann/Maasch (\ref{e:SM91full}) and (b) Palliard/Parrenin (\ref{e:PP04}) models for the more realistic QPn $35$-mode Milankovitch forcing (\ref{e:F2forcing}) at amplitude $k_1$ in obliquity and $k_2$ in precessional components. Compared to Figures~\ref{f:SMLEscanQP}(c) and \ref{f:PPLEscanQP}(c) we find similar features but at a different amplitude of forcing.}
	\label{f:SMPPLEscan}
\end{figure}

Finally, in Figure~\ref{f:SMPPLEscan}, we illustrate scans of LLE for (a) SM91 and (b) PP04  with Milankovitch forcing (\ref{eq:F2oF2p}) with varying obliquity and precessional amplitudes $k_1$ and $k_2$. Observe a similar distribution of regions of locking as in Figures~\ref{f:SMLEscanQP} and \ref{f:PPLEscanQP}, but, as expected, with a considerably more complex bifurcation structure associated with the QPn forcing. Note that the forcing in \cite{DeSaedeleer:2013dk} corresponds to a point on a line $(k_1,k_2)=(k\mu_1,k\mu_2)\approx k(26,76)$.

\section{Discussion}

\label{sec:discuss}

In summary, a lack of chaotic response to forcing is a property of some nonlinear oscillators that have specific structures - in particular, for the van der Pol oscillator in the relaxation oscillation regime, the chaos is associated with canard solutions which only appear as part of the attractor for narrow regions of phase space. Phase oscillators can also avoid chaotic responses; for example, consider the phase oscillator model of \cite{Mitsui:2015hea}:
\begin{align}
\frac{d}{dt} \phi &= \beta+\alpha(\cos \phi+\delta \cos 2\phi)[1+\gamma \Lambda(t)]
\label{e:PO}
\end{align}
where $\phi\in \R/2\pi\Z$ is a periodic phase variable. The one-dimensional nature of the system means that LLEs on attractors are non-positive: chaos is not possible, even though strange non-chaotic attractors (SNAs) can appear in this system \cite{Mitsui:2015hea}.

We show here that more general oscillator models, including some physically motivated ones, can show larger regions in parameter space where there are chaotic attractors. We find this for the van der Pol-Duffing oscillator as well as for the models of \cite{Saltzman:1991jl} and \cite{Paillard:2004dn}, and this is the case not only for periodic forcing, but also for two-frequency forcing and a realistic approximation of Milankovitch forcing. We expect to find similar behaviour even in systems that have no stable limit cycle in the unforced case, but if there is non-linear resonance with the forcing \citep{Marchionne:2016vo}.

\begin{figure}
\centering
		\includegraphics[width=0.48\textwidth]{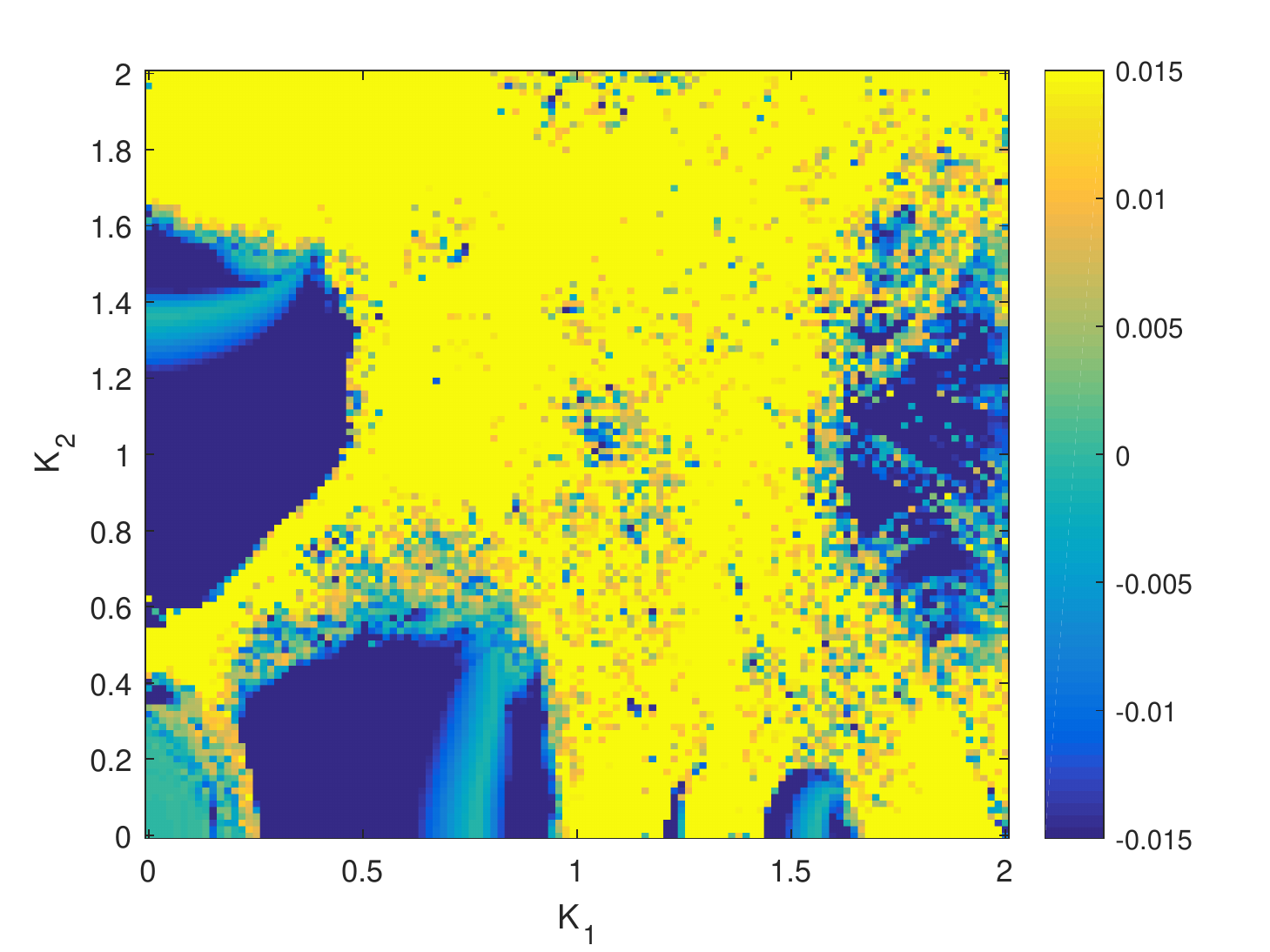}~
		\includegraphics[width=0.48\textwidth]{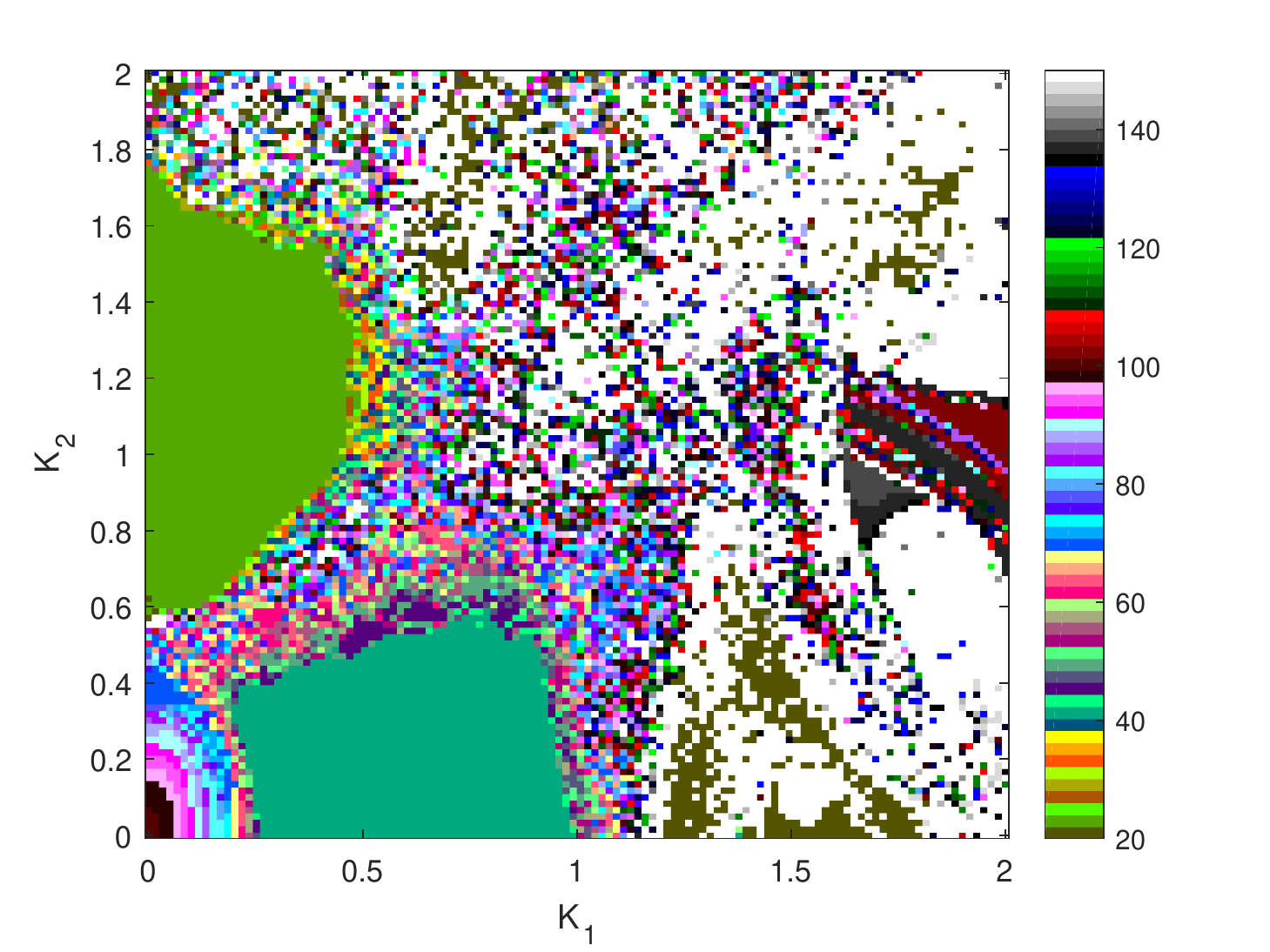}

		(a)\hspace{5cm}(b)
\caption{Lin and Young oscillator (\ref{e:LY}) with QP2 forcing (\ref{e:F1forcing}): (a) shows LLE while (b) the average period. Observe the large yellow regions of chaotic behaviour in (a). \label{f:LYscan} }
\end{figure}

Certain features need to be present to show prevalent chaotic response, firstly the forcing needs to be of sufficient amplitude and, at least for two dimensional oscillators, there needs to be some shear in phase space near the limit cycle. This has been found to give rise to chaotic response in a variety of systems in that it gives a mechanism for the required stretch-and-fold in phase space \citep{Blackbeard:2011}. To illustrate, consider the forced shear oscillator of  \cite{LinYoung:2008}:
\begin{align}
\frac{d}{dt}\theta & = \frac{2\pi}{P}+\sigma y\nonumber\\
\frac{d}{dt}y & = -\lambda y + \gamma\sin(\theta)\Lambda(t)
\label{e:LY}
\end{align}
where $\theta\in \R/(2\pi\Z)$ is a phase variable and $y\in\R$. If we choose parameters $P=100$, $\sigma=3$, $\lambda=0.2$, $\gamma=0.1$ and scan through $k_1$ and $k_2$ we find  large regions of strongly chaotic responses for QP2 forcing (\ref{e:F2forcing}); see Figure~\ref{f:LYscan}. 

\subsection{Implications for modelling the Pleistocene ice ages}

This study does not imply either the presence or absence of a chaotic response to forcing of the physical `ice-age' oscillator by astronomical forcing. However, it does suggest that a chaotic response cannot be ruled out simply because it is rare for the forced van der Pol oscillator. Indeed, both of the physics-based models show more chaos more readily, but still only in regions of parameter space that may or may not be relevant to the late Pleistocene ice ages.

We do note however that in order to get an approximately 100~kyr response to a periodic forcing at approximate 41~kyr period, we need to be in a region of moderate forcing amplitude - too great and we find 1:1 locking, too little and we find no locking. This happens to also be a region where we can find regions of chaotic response for the PP04 and SM91 models.

If there is a chaotic response with LLE $\lambda>0$, this would place a fundamental limit to the predictability of future ice ages, and in particular the onset of rapid deglaciations. Roughly speaking, suppose an estimation of the error in initial fraction of ice for an accurate ice age model is $0<\varepsilon<1$. This will grow at approximate rate $\lambda$, meaning there is effectively a time horizon $T_h$ such that $\varepsilon e^{\lambda T_h}=O(1)$, i.e.  $T_h=-\ln \varepsilon/\lambda$. Since convergence of Lyapunov exponents is highly non-uniform, a more sophisticated estimate along the lines discussed in \cite{vonderHeydt:2017dw} which includes state dependence of the convergence could imply the existence of longer or shorter time horizons depending on current state. For the examples considered in Figure~\ref{f:LLEexample}, the implied LLE is approximately $0.004$ corresponding to a timescale of $e$-folding of errors that is $\lambda^{-1}\approx 250$kyr. As for weather forecasting, the time horizons represent a limit beyond which detailed forecasts are likely to be no better than average state estimation. 


We remark on a number of possible interesting directions for further work. One is clearly to work with more realistic and accurate models of the ice-age oscillations. Another is to improve the forcing - the Milankovitch 65$^o$N summer solstice forcing is clearly only an idealization. However there are still considerable challenges to use data to constrain the models in a meaningful way, in particular to determine whether a model with positive LLE or negative LLE is more appropriate, and if negative whether any response is an SNA or not. Indeed, it would be interesting to explore the appearance of SNAs that may appear during the breakdown to chaos \citep{FeudelKuznetsovPikovsky:2006} for these systems.


\subsection*{Acknowledgements}

The authors thank the Past Earth Network (EPSRC grant number EP/M008363/1) and ReCoVER (EPSRC grant number EP/M008495/1) for partial support of visits of CDC and AvdH to the University of Exeter. We especially thank Michel Crucifix, Peter Ditlevsen and Sebastian Wieczorek for their inspiring and interesting comments, and we thank Trevor Bailey for the use of his office.

%
%


\newpage

\appendix

\section{Computation of largest Lyapunov exponents}
\label{sec:appendix}

\begin{figure}
	\begin{center}
		\includegraphics[width=12cm]{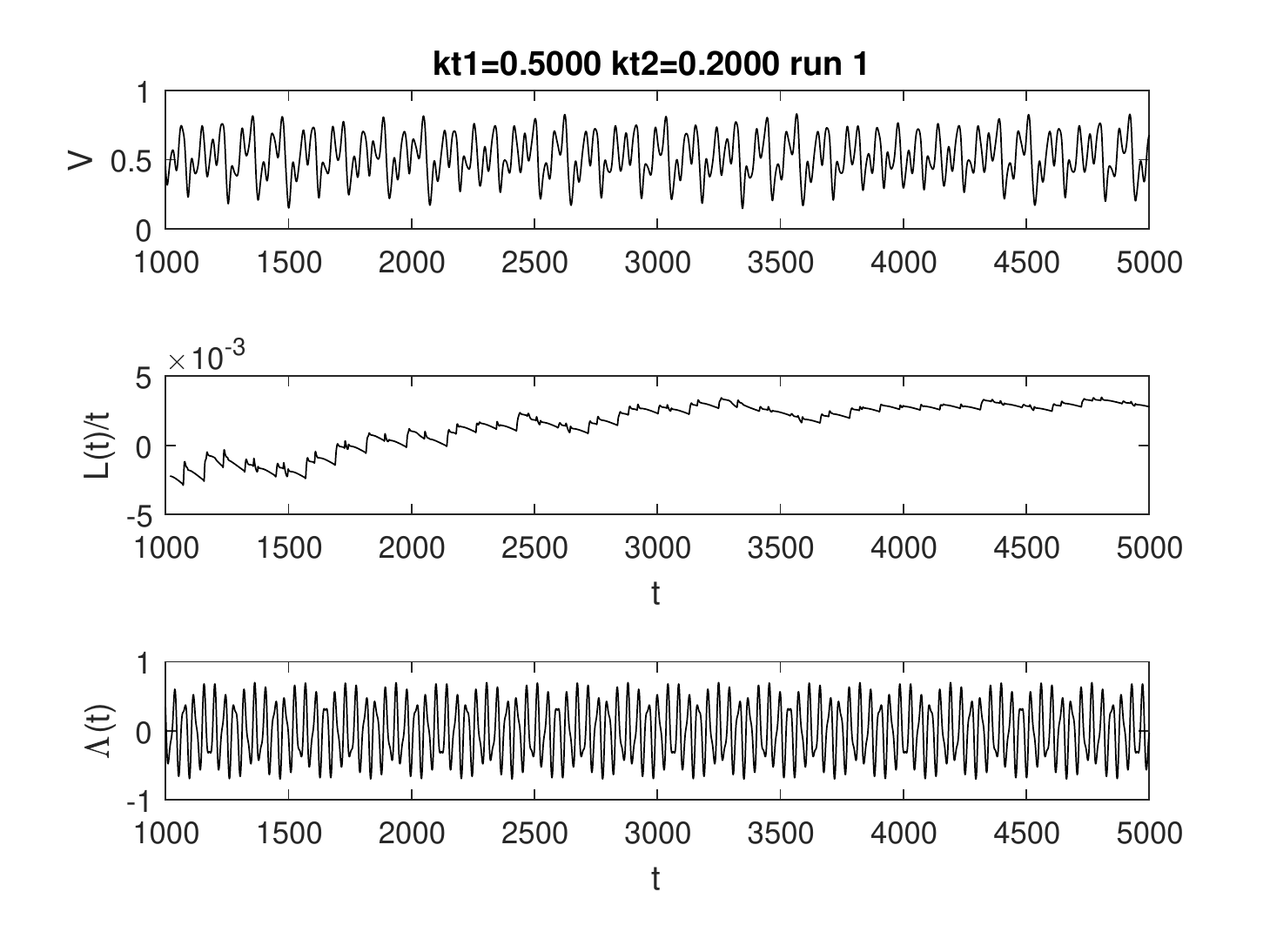}		
	\end{center}
\caption{
The response $V(t)$ (top panel) for a random initial condition of the Paillard-Parrenin model (\ref{e:PP04}) as in Figure~\ref{f:PPTS}(c) for QP2 forcing $\Lambda(t)$ (bottom panel). The middle panel shows the quantity $L(t)/t$ where $L(t)$ is computed using (\ref{e:modvariational}). Observe convergence to a LLE which, in this case, is positive.
\label{f:LLEexample}}
\end{figure}

We compute the largest Lyapunov exponent (LLE) using the variational equations (\ref{e:variational}) with a logarithmic radial variable, by computing trajectories of the augmented system
\ba
 \frac{d}{dt}x & = & f(x,\Lambda(t))\nonumber\\
 \frac{d}{dt}w & = & W(t,w)w\label{e:modvariational}\\
 \frac{d}{dt}L & = & V(t,w)\nonumber
\ea
where $x(t)\in\R^d$, $0\neq w(t)\in\R^d$, $L(t)\in\R$. In particular we define
\be
V(t,w)=\frac{\langle w, Df(x(t),\Lambda(t))w\rangle}{\langle w, w\rangle}\in\R
\label{eq:Wdef}
\ee
and
\be
W(t,w)=Df(x(t),\Lambda(t))-V(t,w)I_d\in\R^{d\times d}
\label{eq:Vdef}
\ee
where $I_d$ is the $d\times d$ identity matrix.
From (\ref{e:modvariational},\ref{eq:Wdef},\ref{eq:Vdef}), it follows that 
\begin{eqnarray*}
\frac{d}{dt}\|w(t)\|^2 & = & 2\langle w(t),W(t,w)w\rangle\\
& = & 2\langle w(t),[Df(x(t),\Lambda(t))-V(t,w)I_d]w(t)\rangle\\
& = & 2\langle w(t),[Df(x(t),\Lambda(t))-\frac{\langle w, Df(x(t),\Lambda(t))w\rangle}{\langle w, w\rangle}]w(t)\rangle\\
& = & 2\langle w(t),Df(x(t),\Lambda(t)) w(t)\rangle \left[1-\frac{\|w(t)\|^2}{\|w(t)\|^2}\right]=0
\end{eqnarray*}
and so $\|w(t)\|$ is constant in time $t$. Note that if $v(t)=w(t)e^{L(t)}$ where $w(t)$, $L(t)$ are solutions of (\ref{e:modvariational}), it follows that
\begin{eqnarray*}
\frac{d}{dt}v(t) &=& \frac{d}{dt}[w(t)] e^{L(t)}+w(t) \frac{d}{dt}[e^{L(t)}]
\\
&=& W(t,w)w(t) e^{L(t)}+V(t,w) w(t)e^{L(t)}\\
&=& [W(t,w)+V(t,w)]v(t)\\
&=& Df(x(t),\Lambda(t))v(t)
\end{eqnarray*}
Hence $v(t)$ is a solution of (\ref{e:variational}). This means that for typical $w(0)\neq 0$ and $x(0)$ approaching the attractor, the LLE can be found as the rate of linear growth of $L(t)$:
$$
\lambda_1=\lim_{t\rightarrow \infty} \frac{1}{t}L(t).
$$
In practice, we compute $\lambda_1$ for fixed, large $t$ and  choose $L(0)=0$. As an example, Figure~\ref{f:LLEexample} shows a numerical approximation of the LLE for the PP04 model (\ref{e:PP04}) as in Figure~\ref{f:SMTS}(f): observe apparent convergence of $L(t)/t$ to a positive LLE.

\end{document}